\documentclass[11pt]{article}
\usepackage{amsmath}
\usepackage{epsfig,amssymb,latexsym}
\usepackage[all]{xy}

\usepackage[colorlinks=true, urlcolor=blue, pdfborder={0 0 0}]{hyperref}
\usepackage{amsmath}%
\usepackage{amsfonts}%
\usepackage{amssymb}%
\usepackage{graphicx}
\usepackage{subcaption}
\usepackage{bbm} 
\usepackage{multirow}
\usepackage[top=1.5in, bottom=1.5in, left=1.3in, right=1.2in]{geometry}

\numberwithin{equation}{section}

\def\d{\mathrm{d}}

\newcommand{\zs}[1]{{\mathchoice{#1}{#1}{\lower.25ex\hbox{$\scriptstyle#1$}}
{\lower0.25ex\hbox{$\scriptscriptstyle#1$}}}}

\newcommand{\be}{\begin{equation}}
\newcommand{\ee}{\end{equation}}
\newcommand{\bea}{\begin{eqnarray}}
\newcommand{\eea}{\end{eqnarray}}
\newcommand{\beas}{\begin{eqnarray*}}
\newcommand{\eeas}{\end{eqnarray*}}

\newtheorem{theorem}{Theorem}[section]

\newtheorem{assumption}{Assumption}

\newtheorem{corollary}[theorem]{Corollary}

\newtheorem{lemma}[theorem]{Lemma}

\newtheorem{proposition}[theorem]{Proposition}

\numberwithin{equation}{section}

\newenvironment{proof}[1]{\addvspace{\medskipamount}\par\noindent{\it Proof#1}.}
{\unskip\nobreak\hfill$\Box$\par\addvspace{\medskipamount}}

\newcommand{\EE}{\mathbb{E}}

\def\F{\mathcal{F}}

\def\R{\mathbb{R}}

\newcommand{\ree}{\mathbb{R}}

\newcommand{\PP}{\mathbb{P}}
\newcommand{\QQ}{\mathbb{Q}}
\newcommand{\EX}{\mathbb{E}}

\newcommand{\esssup}{\text{ess sup}}
\newcommand{\wh}{\widehat}

\newcommand{\wt}{\widetilde}

\newcommand\cG{{\cal G}}
\newcommand\cF{{\cal F}}

\def\proof{{\noindent \bf Proof. }}

\def\bbr{{\mathbb R}}

\begin{document}

\title{Non-concave expected utility optimization with uncertain time horizon}

\author{Christian Dehm\footnote{Institute of Insurance Science and Institute of Financial Mathematics, University of Ulm, Ulm, Germany, christian.dehm@uni-ulm.de.}\qquad\qquad Thai Nguyen\footnote{\'{E}cole d'Actuariat, Universit\'{e} Laval,  Qu\'ebec city, Canada, thai.nguyen@act.ulaval.ca.} \qquad\qquad Mitja Stadje\footnote{{Institute of Insurance Science and Institute of Financial Mathematics, University of Ulm, Ulm, Germany}, mitja.stadje@uni-ulm.de
\newline
{\em Keywords and phrases.} non-concave utility, uncertain time horizon, porftolio selection.
\newline
{\em  AMS Classification.} 49N99, 91G80, 91G10, 93E20.
}
}

\date{}

\maketitle
\begin{abstract} 
We consider an expected utility maximization problem where the utility function is not necessarily concave and the time horizon is uncertain. We establish a necessary and sufficient condition for the optimality for general non-concave utility function in a complete financial market. We show that the general concavification approach of the utility function to deal with non-concavity, while being still applicable when the time horizon is a stopping time with respect to the financial market filtration, leads to sub-optimality when the time horizon is independent of the financial risk, and hence can not be directly applied. For the latter case, we suggest a recursive procedure which is based on the dynamic programming principle.  We illustrate our findings by carrying out a multi-period numerical analysis for optimal investment problem under a convex option compensation scheme with random time horizon. We observe that the distribution of the non-concave portfolio in both certain and uncertain random time horizon is right-skewed with a long right tail, indicating that the investor expects frequent small losses and a few large gains from the investment. While the (certain) average time horizon portfolio at a premature stopping date is unimodal, the random time horizon portfolio is multimodal distributed which provides the investor a certain flexibility of switching between the local maximizers, depending on the market performance. The multimodal structure with multiple peaks of different heights can be explained by the concavification procedure, whereas the distribution of the time horizon has significant impact on the amplitude between the modes.

\end{abstract}

\section{Introduction}

A classical problem in optimal control theory and mathematical finance is to maximize the expected reward or utility over all admissible terminal positions (portfolios) starting with an initial investment in the time horizon $[0,T]$, 
where $T>0$ is given upfront and 
the objective (utility) function is a concave.  Such a utility maximization problem in a continuous-time setting dates back to Merton \cite{Merton1971} with the underlying stochastic processes representing a financial market. Merton's pioneering work has been extended in several directions e.g. by assuming more general structures of preferences, by incorporating additional  randomness to the underlying risk processes, or by including a risk constraint to the optimization problem, see among many others e.g. Biagini \cite{Biagini_survey}, Wong et al. \cite{Wong}, or Karatzas et al. \cite{Karatzas} for a broad discussion.  


\vspace{1mm}

In this work, we investigate an extension of the Merton problem to the case where the utility function is not necessarily concave and the time horizon is random. 

Let us briefly mention some of the most relevant literature. Most optimal control-type problems have a fixed known time horizon. However, in reality such a natural fixed maturity does not exist and instead exogenous or endogenous events determine the end of the optimal control/optimal investment problem. An early paper by Yaari \cite{Yaari} looks at the investment problem of an individual with an uncertain time of death in a simplified case with purely deterministic investment opportunities. Yaari's paper is extended to discrete-time settings with multiple risky assets. 
Optimal life-cycle consumption and investment is studied by Merton \cite{Merton1971}, where the time horizon uncertainty is reflected by the first jump of an independent Poisson process with constant intensity. Richard \cite{Richard} solves in closed-form an optimal portfolio choice problem with an uncertain time of death and the presence of life insurance. 
In these works, the time horizon uncertainty can be treated as additional discount factor and closed-form solution can be provided by using dynamic programming principle for concave utility functions. A more complete setting for concave utility maximization with a continuous time horizon distribution in a complete financial market has been studied in Blanchet et al. \cite{Blanchet}. Bouchard and Pham \cite{BouchardPham} investigate a concave utility maximization in an incomplete market with general uncertain time horizon structure. All the mentioned works leave the case where the objective utility is not necessarily concave e.g. \cite{Carpenter} as an open problem.  To the best of our knowledge, the non-concave utility maximization problem under random time horizon has not yet been investigated.

The literature of non-concave optimization with certain time horizon is vast, see for instance Aumann and Perles \cite{AumannPerles}, Basak and Makarov \cite{BasakMakarov}, Bensoussan et al. \cite{Bensoussan}, Bichuch and Sturm \cite{BichuchSturm}, Carassus and Pham \cite{CarassusPham}, Carpenter \cite{Carpenter}, Chen et al. \cite{Chen},   
Larsen \cite{Larsen}, Reichlin \cite{Reichlin}, Rieger \cite{Rieger2012} and Ross\cite{ross2004}. For non-concave optimization with constraints see Nguyen and Stadje \cite{Stadje} or Dai et al. \cite{Dai}. In these works in the finance and the OR literature the non-concavity arises typically from non-linear, option-type managerial compensations. Such remuneration schemes have been seen in industry as one way to overcome potential principal-agent issues and are supposed to align the incentives of managers with the ones of owners.

Another important application of non-concave investment problem relate to participating insurance contracts which have been extensively used in European and non-European life insurance markets, see the references provided in the introduction. Typically, to buy a participating insurance policy, the policyholder pays a lump sum premium upfront and the capital saved is invested in a self-financing way, subject to annual interest, where the insurance company offers a (minimal) guarantee. An example is given by so called ``flexibility rider contract'' which have gained popularity recently due to the current low interest rate development where the decision variable is the riskiness of the investment pool, see
\cite{Chen} and the references within. 

In positive economic developments, the policyholder receives a surplus, while in case of bad economic developments, the insurance company carries the loss. 
Hence, a participating insurance contract may be regarded as an option-type financial instrument, leading to a non-concave utility function. In such a context our work to the best of our knowledge is the first one which is able to include the randomness of the lifetime into the investment problem (instead of simply assuming a fixed pre-specified time-horizon).    


As we aim to obtain some explicit result in the illustration section, we extensively consider the option compensation problem in \cite{Carpenter} where the utility function admits only one concavification interval but with a random time horizon which has a discrete distribution on the universal time interval $[0,T]$. We remark that our results can be extended to settings with a continuous distribution time horizon.

Our contribution is fourth-fold. First, we show that when the time horizon is a stopping time with respect to the financial market filtration, the general approach of concavificiation techniques as described in \cite{Rieger2012} to deal with non-concavity can be applied. This is an extension of the result in \cite{Carpenter}, complementing the result in \cite{BouchardPham} (Proposition 4.3) to random time horizons in complete markets. Second, when $\tau$ is independent of the financial risk and the market is therefore incomplete, we establish necessary and sufficient conditions for the optimality for general utility functions. Third, also 
for the case where $\tau$ is independent of the financial risk, we show that optimizing the concavified version of the utility function will lead to sub-optimality with a potentially significant expected utility loss and suggest a recursive procedure which is based on the dynamic programming principle to solve the optimization problem in this situation. Fourth, we illustrate our finding by carrying out a multiple period numerical analysis for the non-concave option compensation problem with random time horizon thoroughly exploring the effect of randomness on managerial compensation schemes and participing insurance contracts. This is computationally challenging because the optimal multiplier obtained by the concavified problem in one period is a random variable that depends on the market realizations at the end of the previous period. 

We numerically show that under an uncertain time horizon which imposes a new randomness that cannot be fully hedged by only using the available financial instruments, the concavified problem strategy is super-optimal and leads to an expected utility loss. In addition, due to concavification, the distribution of the wealth at exiting times of the non-concave optimization problems is right-skewed with a long right tail, indicating that the investor can expect frequent small losses and a few large gains from the investment. Intuitively, a positively skewed distribution of investment returns is generally desirable by the agent with option-liked compensation payoff because there is some probability to gain huge profits that can cover all the frequent small losses. Under the premature exiting risk, the wealth at an exiting time exhibits a bimodal distribution with peaks of different heights. The bimodal structure can be explained by the concavification procedure whereas the distribution of the exiting time $\tau$ has significant impact on the  amplitude between the two modes. When the concavified utility at an exiting time is affine in many open intervals, the corresponding wealth is intuitively expected to be of multimodal distribution.

The remainder of the paper is organized as follows: First, we describe a specific complete financial market setting and introduce the uncertain investment time in Section \ref{ch:economy}. We present our necessary and sufficient condition for optimality for non-concave general utility functions in Section \ref{Se:Nonconcavegeneral}. We show that the concavification technique is not applicable in a non-concave setting with random time horizon which induces additional risk to the financial market, and derive a dynamic programming principle for such a non-concave optimization with uncertain time horizon in Section \ref{Se:DPP}. In Section \ref{ch:examples}, we investigate the case of power utility and perform a numerical study for non-concave optimization with time horizon uncertainty. We study the case when the time horizon is a stopping time with respect to the financial market filtration in Section \ref{Sec:stopping}. Finally, Section \ref{Se:conclusion} summarizes our main results.

\section{Financial market and the optimal investment problem} \label{ch:economy} \noindent
Let $[0,T]$ with $0 < T < \infty$ be the maximal time span of the economy and $W$ is  $n$-dimensional Brownian motion in a probability space $(\Omega, \mathcal{A}, \mathbb{P})$. 

\subsection{The financial market}
For the market setup, we assume that the prices of $n$ risky assets $S$ are modelled as a geometric Brownian motion, i.e.,
\begin{align*}
\frac{dS_t^{i}}{S_t^{i}} = \mu^{i}_t dt + \sum_{j=1}^{n} \sigma^{i,j}_t dW^{j}_t,~ i = 1, \dots, n,
\end{align*}
where the superscript $i$ denotes the $i$-th entry of the corresponding vector or $(i,j)$ the entry in the $i$-th row and $j$-th colomn of a matrix and we use the subscript $t$ to denote the time index $t$. We use the notation $\mu = (\mu^{i})_{1 \leq i \leq n}$ and $\sigma = (\sigma^{i,j})_{1 \leq i,j \leq n}$ for the corresponding vector or matrix, respectively. Additionally to these risky assets, we consider a \textit{bond} $B$, given by $dB_t = B_t r_t dt$,
where $r$ denotes the (deterministic) interest rate. The information in the market is captured by the augmented filtration $\mathcal{F} =(\mathcal{F}_t)_{t \geq 0}$ generated by the Brownian motion, satisfying the usual conditions and $\mathcal{F}_{0}$ is trivial. We assume that the coefficients $\mu$, $r\geq 0$ are bounded and deterministic and the volatility $\sigma$ is bounded, deterministic, invertible with bounded inverse $\sigma^{-1}$.

In this arbitrage-free financial market, there exists a unique equivalent martingale measure $\QQ$ with Radon-Nikodym density $M$ as the solution of $dM_t = - M_t \theta_t dW_t$, where $\theta_t := \sigma^{-1}_t (\mu_t - r_t \textbf{1})$. Further we define $\xi_t := \exp \left( - \int_{0}^{t} r_s ds \right) M_t$. By It\^{o}'s formula 
$
d\xi_t 
= - \xi_t r_t dt - \xi_t \theta_t dW_t,
$
and
$$
\xi_t =  \exp \left( - \int_{0}^{t} (r_s + \frac{1}{2} \theta^{T}_s \theta_s) ds - \int_{0}^{t} \theta_s dW_s \right). 
$$
We consider the economy in the usual frictionless setting, where stocks and bonds are infinitely divisible and there are no market frictions, no transaction costs etc. 
Additional to the financial market setting, we consider a random time-horizon $\tau$, where $\tau$ is a positive discrete random variable independent of $\mathcal{F}$. In particular, $\tau$ is not an $\mathcal{F}$-stopping time. 
Let $\mathcal{F}^{\tau}=(\mathcal{F}_t^{\tau})_{0\leq t\leq T}$ with $\mathcal{F}_t^\tau$ being the $\sigma$-algebra generated by $({\bf 1}_{\tau\le s})_{0\leq s\le t}$. Define $\mathcal{G}=\mathcal{F}\vee\mathcal{F}^{\tau}$. $\QQ$ can be extended to $\mathcal{G}_T$ by defining $\QQ(A):=\mathbb{E}[\frac{d\QQ}{d\mathbb{P}}{\bf 1}_A]$ for any $A\in\mathcal{G}_T$. We note that any $\mathcal{G}$-martingale is also an $\mathcal{F}$-martingale. 



\subsection{Utility function} \label{ch:utility_function} 

In the sequel
we consider a general not necessarily concave utility function {$U : [0,\infty) \to \ree$ } which is non-constant, increasing, continuous, has left- and right-hand side derivatives and satisfies the growth condition
\begin{align} \label{eq:utility_growth}
\lim_{x \to \infty} \frac{U(x)}{x} = 0.
\end{align}
We set $U(x) = - \infty$ for $x < 0$ to avoid ambiguity and define 
$U(\infty) := \lim_{x \to \infty} U(x)$.
%
We do \emph{not} assume that $U$ is concave or strictly increasing.
In a concave setting, Equation \eqref{eq:utility_growth} is equivalent to  $U'(\infty) = 0$, which is part of the Inada condition. We note that %
Equation \eqref{eq:utility_growth} and the assumption $ U(\infty) > 0$ imply that there exists a concave function ${U}^c: \ree \to \ree\cup \{ - \infty \}$ that dominates $U$, i.e., ${U}^c \geq U$.
The following result explains why we can consider $\mathcal{F}$-predictable, instead of  $\mathcal{G}$-predictable investment strategies.
\begin{lemma}\label{Le:21}
Suppose that $\pi_s\in\mathcal{L}^2_{loc}(d\PP,(\mathcal{G}_s)_{0\leq s\leq T})$. Then there exists a strategy $\overset{\sim}{\pi}_s\in\mathcal{L}^2(d\PP,(\mathcal{F}_s)_{0\leq s\leq T})$:
\begin{equation*}
\int_0^{\tau\wedge T}\pi_s \sigma_s dW_s^\QQ=\int_0^{\tau\wedge T}\overset{\sim}{\pi}_s \sigma_s dW_s^\QQ.
\end{equation*}
\end{lemma}
\proof Denote by $Y_s=\pi_s\sigma_s$ for $0\leq s\leq T$. Prop. 2.11 in \cite{filtration} yields that the $\cG$-predictable process $Y$ can be expressed as 
$Y=y{\bf 1}_{[0,\tau]}+g(\tau){\bf 1}_{]\tau,T]}$
where $(y_s)_{0\leq s\leq T}$ is $\cF$-predictable and $g_t(\omega ,u);t\geq u$ is a $\mathcal{P}\otimes\mathcal{B}([0,T])$ random function. Set $\tilde{\pi}_s=y_s\sigma_s^{-1}$. Then $$\pi_s\sigma_s=Y_s=y_s=\tilde{\pi}_s\sigma_s\quad \mbox{on }0\leq s<\tau\wedge T.$$ This entails that $\int_0^{\tau\wedge T}\pi_s\sigma_s dW_s^\QQ=\int_0^{\tau\wedge T}\tilde{\pi}_s\sigma_s dW_s^\QQ$ and the lemma follows.\endproof

\subsection{Admissible strategy} \label{ch:admissible_strategy} 
We consider an investor 
putting the amount $\pi^i_s$ in each risky asset at time $0 \leq s \leq T$. By considering a self-financing portfolio, the amount $P_s - \sum_{i=1}^{n} \pi^{i}_s$ is invested in the bond. 
We use the notation $(P_s^{t,\pi,x}, t \leq s \leq T)$ for the wealth process at time $s$, developed from an initial capital\\ $P_t^{\pi} := P_t^{t,\pi,x} = x>0$ at time $t$ under a self-financing strategy $\pi$ where $\pi^i$ denotes the amount invested in asset $i$. Then $P_s^{t,\pi,x}$ evolves according to the stochastic differential equation
\begin{align} \label{eq:dP}
dP_s^{t,\pi,x} = P_s^{t,\pi,x} r_s ds + \pi_s \left[ ( \mu_s - \textbf{1}r_s) ds + \sigma_s dW_s \right].
\end{align}
We call a portfolio $(\pi_t ,~ 0 \leq t \leq T)$ \textit{admissible}, if $\pi$ is progressively measurable w.r.t. $\cF$, locally square-integrable, i.e., 
$ \sum_{i = 1}^{n} \int_{0}^{T} (\pi^i_s)^2 ds  < \infty$ a.s., and the associated wealth process is non-negative. 
By Girsanov's theorem as long as $\pi$ is locally square-integrable $(\xi_sP_s^{t,\pi,x})_s$ is always a local martingale. For the set of admissible portfolios with initial capital $x$ at time $t$, we use the notation
\begin{align} \label{eq:def_admissible}
&{\Pi}(t,x) := \left\{ \pi_s: t \leq s \leq T, ~P_t^{t,\pi,x} = x,~ \pi \text{ is admissible and } P_s^{t,\pi,x} \geq 0 \right\}.
\end{align}
We define $\tilde{P}_t := P_t \exp \left( - \int_{0}^{t} r_s ds \right)$ as the \textit{discounted portfolio process}.

\section{Non-concave optimization problem with random time horizon}\label{Se:Nonconcavegeneral}
Assume that the agent evaluates his investment performance at times $0=:T_0< T_1 < T_2 < \dots < T_n:=T$ with respect to the weights $p_i:=\PP(\tau=T_i)$, $i=1,\cdots,n-1$, and $p_n=\PP [\tau \geqslant T_n]$, with $\sum_{i=1}^{n} p_i =1$. Let $\wt{\Pi}$ be the set of all portfolios in $\pi$ that are progressively measurable with respect to $\cG$, locally square-integrable with non-negative associated wealth process.
In our complete financial market setup, we consider the problem 
\begin{align} 
V_{\tau}(x,U) &=\sup_{\pi \in\wt{\Pi}(0,x)} \EX \left[U(P_{T\wedge\tau} ) \right]= \sup_{\pi \in{\Pi}(0,x)} \EX \left[U(P_{T\wedge\tau} ) \right]=\sup_{\pi \in \Pi(0,x)} \EX \left[\sum_{i=1}^{n} p_i U(P_{T_i} ) \right],\label{eq:general_problem}
\end{align}
where the second equality holds by Lemma \ref{Le:21}. 
Define 
\begin{align} \label{eq:def_Cx}
C_\tau(x) := \bigg\{
&P=(P_{0},\cdots,P_{n}): P_i\ge 0,\, \mathcal{F}_{T_i}\text{-measurable with} \,\tilde{P}_{i} =:e^{-r T_i} P_i= \tilde{P}_{ i-1}  + \int_{ T_{i-1}}^{t} \sigma_s \pi_s^i dW_s^{\QQ}\nonumber \\  &
\quad
 \mbox{ and }t\in (T_{i-1},T_i],
\quad \pi^{i}\in \Pi(T_{i-1},P_{i-1}), \quad i=1,\cdots,n
\bigg\}, 
\end{align}
where by convention $P_0=x$. Note that for $P=(P_{0},\cdots,P_{n}) \in C_\tau(x)$, we have $
\EX \left[ \sum_{i=1}^{n} p_i \xi_{T_i} P_{i} \right] \leq x.$
Furthermore, it is clear that the strategy $$\pi_t^{(n)}(P_{0},\cdots,P_{n}) =\sum_{i=1}^{n}\pi_t^i{\bf 1}_{(T_{i-1},T_{i}]}(t)\in \Pi(0,x)$$
is locally square integrable. Denote by  $P^{x,\pi^{(n)}}:=P^{x,\pi^{(n)}(P_{0},\cdots,P_{n})}$ the corresponding wealth process (which is equal to $P_i$ at time $T_i$ for $i=0,\ldots,n$). Note that $\xi P^{x,\pi^{(n)}}=\xi P^{x,\pi^{(n)}(P_{0},\cdots,P_{n})}$ is a non-negative local martingale (hence a supermartingale). We say that the supermartingale $\xi P^{x,\pi^{(n)}}$ is generated by the $n+1$-tuple $P=(P_{0},\cdots,P_{n}) \in C_\tau(x)$.
The optimization problem \eqref{eq:general_problem} can be restated in the following way 
\begin{align}
V_{\tau}(x,U) = \sup_{P \in C_\tau(x)} \EX \left[\sum_{i=1}^{n} p_i U(P_i ) \right].\label{eq:manyTprob}
\end{align}
Define 
\begin{equation}
I(a):=\inf\bigg\{y>0|U(y)-ya=\underset{p\ge 0}{\sup}\{U(p)-pa\}\bigg\}.
\label{eq:Iinverse}
\end{equation}
By continuity the supremum and infimum are attained so that $I(x)$ is the smallest $\arg\max$ of the function $y\rightarrow U(y)-yx$. 
Below, $U'$ denotes the right-hand side derivative of $U$. The following result provides a sufficient condition for optimality of the optimization problem \eqref{eq:manyTprob}.
\noindent
\begin{theorem} \label{thm:adapted_process}Let $x>0$. Suppose that there is an adapted process $\nu\geq 0$ with $\nu_0=U'(x)$ such that the process $\xi P^{x,\pi^{(n)}(x,I(\nu_{T_1} \xi_{T_1}),\cdots, I(\nu_{T_1}\xi_{T_n}))}$ generated by the $n+1$-tuple $P^{*}:=(x,I(\nu_{T_1} \xi_{T_1}),\cdots, I(\nu_{T_n}\xi_{T_n}))$ is a martingale and {$ \sum_{i=1}^{n} p_i \nu_{T_i} $} is a constant.
Then,  $P^{*}$ solves the optimization problem \eqref{eq:manyTprob}.
\end{theorem} 
\proof
Let $\sum_{i=1}^{n} p_i \nu_{T_i}=y$ which is a constant by assumption. Then for any for any $Y=(x,Y_{T_1},\cdots, Y_{T_n})\in C_\tau(x)$ by construction the process $\xi P^{x,\pi^{(n)}(x_0,Y_{T_1},\cdots, Y_{T_n})}$ is a non-negative local martingale. Let $\tau_m$ be the corresponding localizing sequence. Then 
\begin{align*}
\EX[\sum_{i=1}^{n} p_i \nu_{T_i} \xi_{T_i\wedge \tau_m} Y_{T_i\wedge \tau_m} ]
&=\sum_{i=1}^{n} \EX\bigg[p_i\nu_{T_i}\EX[  \xi_{ T\wedge \tau_m} Y_{T\wedge \tau_m}\vert \cF_{T_i}]\bigg]\nonumber\\
&=\sum_{i=1}^{n}\EX\bigg[ p_i\nu_{T_i}\xi_{ T\wedge \tau_m} Y_{T\wedge \tau_m}\bigg]=\EX\bigg[(\sum_{i=1}^{n} p_i\nu_{T_i})\xi_{ T\wedge \tau_m} Y_{ T\wedge \tau_m} \bigg]
=x y, 
\end{align*}
Passing to the limit and using Fatou's lemma yields 
\begin{align}
\label{eq:dualityineq}
\EX[\sum_{i=1}^{n} p_i \nu_{T_i} \xi_{T_i} Y_{T_i} ]
\leq xy, \quad \forall \,Y=(x,Y_{T_1},\cdots, Y_{T_n})\in C_\tau(x).
\end{align}

The process $Z:=x^{-1}\xi P^{x,\pi^{(n)}(x,I(\nu_{T_1} \xi_{T_1}),\cdots, I(\nu_{T_1}\xi_{T_n}))}$ defines a density process of a probability measure $\QQ^\nu<<\PP$ as it is a martingale with initial value equal to 1. Due to the construction of $P^{x,\pi^{(n)}(x,I(\nu_{T_1} \xi_{T_1}),\cdots, I(\nu_{T_1}\xi_{T_n}))}$ it can be observed that $Z_{T_i}=x^{-1}\xi_{T_i}I(\nu_{T_i} \xi_{T_i})$. Therefore, we obtain
$$
\EX[\sum_{i=1}^{n} p_i \nu_{T_i} \xi_{T_i}I(\nu_{T_i} \xi_{T_i})]=x\EX^{\QQ^\nu}[\sum_{i=1}^{n} p_i \nu_{T_i}  ]=xy.
$$
For any admissible $Y=(x,Y_{T_1},\cdots, Y_{T_n})\in C_\tau(x)$ we have 
\begin{align*}
\EX \left[\sum_{i=1}^{n} p_i U(I(\nu_{T_i} \xi_{T_i})) \right]&=\EX \left[\sum_{i=1}^{n} p_i U(I(\nu_{T_i} \xi_{T_i})) \right]- x\EX^{\QQ^\nu}[\sum_{i=1}^{n} p_i \nu_{T_i} ]+xy\\
&=\EX \left[\sum_{i=1}^{n} p_i U(I(\nu_{T_i} \xi_{T_i})) \right]- \EX[\sum_{i=1}^{n} p_i \nu_{T_i} \xi_{T_i}I(\nu_{T_i} \xi_{T_i})]+xy\\
&=\EX \left[\sum_{i=1}^{n} p_i \sup_{X\ge 0}\bigg(U(X) -\nu_{T_i} \xi_{T_i}X\bigg)\right ]+xy\\
&\ge \EX \left[\sum_{i=1}^{n} p_i \bigg(U(Y_{T_i}) -\nu_{T_i} \xi_{T_i}Y_{T_i}\bigg) \right ]+xy
\ge\EX \left[\sum_{i=1}^{n} p_i U(Y_{T_i})\right ],
\end{align*}
where we have used \eqref{eq:dualityineq} in the last step. This implies the optimality of $P^{{*}}$.
\endproof


We now seek for a necessary condition for optimality. The following is the main theorem of this section which generalizes results by 
Blanchet et al. \cite{Blanchet} to non-concave settings. 
Let $U'$ be the right-hand side derivative of $U$. We need the following assumption.

\begin{assumption} \label{ass:existence_integrals}
We assume that $( P^{{*}}_{T_1},\cdots, P^{{*}}_{T_n})$ is an optimal solution to Problem \eqref{eq:manyTprob} such that $ \EX \left[\max_i | {U}(P_{T_i}^{{*}})| \right] < \infty$ for some $\delta >0$, that $\xi P^{*}$ is a square integrable martingale (instead of only a local martingale), and that
$\EX [\max_{i}\left[ v' \left( (1-\delta) P^{{*}}_{T_i}  \right)   P^{{*}}_{T_i}\right]] < \infty$ with $v'$ being a decreasing function with $U'\leq v'$ and  $\EX\bigg[\max_{i}\frac{P^{{*}}_{T_i}}{\xi_{T_i}}\int_0^T\xi_s^2\pi_s^2\sigma_s^2ds\bigg]<\infty.$
\end{assumption}


If $U=U^c$ (concave hull) for all $x>x_0$ for an $x_0$ and if $U'$ is bounded on $[0,x_0]$, we can choose $v'(x)=a+U'(x)$. Denote by $M\subset\R_0^+$ the set of all $y\geq 0$ such that $U$ is not differentiable in $y$ and $\{P^{{*}}_{T_i}=y\}$ is a non-zero set for a $i\in\{1,...,n\}$. 
Note that, since $U$ is not defined on the negative half-line, $U$ is in particular not differentiable in $0$. Denote the set of all $\omega$ with $P^{{*}}_{T_i}(\omega)\in M$ for at least one $i$ by $A^c$, and let $A$ be its complement. In other words, the set $A$ contains at most scenarios with portfolio outcomes at a $T_i$ where the utility function is differentiable. 
\begin{theorem} \label{thm:prop2} Assume that $( P^{{*}}_{T_1},\cdots, P^{{*}}_{T_n})$ is an optimal solution to Problem \eqref{eq:manyTprob} which satisfies Assumption \ref{ass:existence_integrals}. Define $\nu_{T_i}:=\xi_{T_i}^{-1}U'(P^{{*}}_{T_i})$ 
	for $i=1,\cdots,n$ .  Then, it holds that the random variable $\sum_{i=1}^{n} p_i \nu_{T_i}
$ is constant (a.s.) on $A$. 
\end{theorem}



\proof If $A$ is a zero-set the theorem is obvious. So assume $\PP(A)>0$. Consider an admissible (non-negative) portfolio $Y$ with terminal value of the form $Y_T=P^{{*}}_T{\bf 1}_{A^c}+Y_T{\bf 1}_A$ such that $\xi Y$ is a martingale. In particular, we consider a portfolio which at time $T$ agrees with $P^{{*}}_T$ on $A^c$. We define for $0 \leq \varepsilon \leq 1$ the functions $\Phi$ and $\chi$ by $\Phi(\varepsilon) := \EX \left[ \chi(\varepsilon) \right]$ and
$ 
\chi(\varepsilon,\omega) :=  \bigg(\sum_{i=1}^{n} U\left( \varepsilon P^{{*}}_{T_i} + (1- \varepsilon) Y_{T_i} \right) p_i\bigg) 
$
We 
denote the right-hand side derivative of a continuous function $f$ by $f'_+$ and the left-hand side derivative by $f'_-$. Of course, in points where the function is differentiable both limits coincide and the ``$+$'' and ``$-$'' may be omitted, respectively.
Note that $\varepsilon P_T^{{*}}+(1-\varepsilon )Y_T$ is admissible so that $\chi$ is integrable. 
 For $\varepsilon>1-\delta$ we have
$(1 - \delta) P^{{*}}_{T_i} <  \varepsilon P^{{*}}_{T_i} + (1- \varepsilon) Y_{T_i}$.
We calculate 
\begin{align*}
\vert \chi^\prime(\varepsilon,\omega)\vert 
\leq & \sum_{i=1}^{n} p_i \left| v' \left( (1-\delta) P^{{*}}_{T_i} \right)  \right| P^{{*}}_{T_i}
\leq  \max_{i} \left| v' \left( (1-\delta) P^{{*}}_{T_i} \right)  \right| P^{{*}}_{T_i}.
\end{align*}
Denote $\text{sign}(x):=+$ if $x\geq 0$ and $\text{sign}(x):=-$ else. Under Assumption \ref{ass:existence_integrals} we obtain 
$$
\Phi^\prime(\varepsilon) = \EX \bigg[ \sum_{i=1}^{n} U'_{\text{sign}(P^{{*}}_{T_i}  - Y_{T_i})} \left( \varepsilon P^{{*}}_{T_i} + (1- \varepsilon) Y_{T_i} \right) \left(P^{{*}}_{T_i}  - Y_{T_i} \right) p_i
\bigg].
$$
We know that the function $\Phi$ attains its maximum at $\varepsilon = 1$, since $P^{{*}}$ is the optimal solution by assumption. 
Hence,
$ 0 \leq  \Phi^\prime_{-}(1). $ 
Thus, 
\begin{align*}
0 &\leq   \EX \bigg[\sum_{i=1}^{n} U_{\text{sign}(P^{{*}}_{T_i}  - Y_{T_i})}'  \left(  P^{{*}}_{T_i}  \right) \xi_{T_i}^{-1} p_i \left(\xi_{T_i} P^{{*}}_{T_i}  - \xi_{T_i} Y_{T_i} \right)
\bigg].
\end{align*}
Integration by parts yields
\begin{align*}
\EX \left[ \sum_{i=1}^{n} U'_{\text{sign}(P^{{*}}_{T_i}  - Y_{T_i})}  \left(  P^{{*}}_{T_i}  \right) \xi_{T_i}^{-1} p_i \left(\xi_{T_i} P^{{*}}_{T_i}  - \xi_{T_i} Y_{T_i} \right)  \right]&= \EX \left[ (P_T^{{*}} - Y_T) \xi_T \sum_{i=1}^{n} U'_{\text{sign}(P^{{*}}_{T_i}  - Y_{T_i})}  \left(  P^{{*}}_{T_i}   \right) \xi_{T_i}^{-1} p_i  \right]\\
& = \EX \left[ (P_T^{{*}} - Y_T) \xi_T \sum_{i=1}^{n} U'  \left(  P^{{*}}_{T_i}   \right) \xi_{T_i}^{-1} p_i {\bf 1}_A \right].
\end{align*}
since by definition $P_T^{{*}}=Y_T$ on $A^c$ and $U(y)$ is differentiable for $y=P^{{*}}_{T_i}$ for each $i$ so that
$U'(P^{{*}}_{T_i})=U'_+(P^{{*}}_{T_i})=U'_-(P^{{*}}_{T_i}).$ Hence,
\begin{align} \label{eq:FOC1}
0 \leq   \EX \left[ \xi_T \left( P^{{*}}_T  -  Y_T \right) \left( \sum_{i=1}^{n} U' \left(  P^{{*}}_{T_i} \right) \xi_{T_i}^{-1}  p_i 
 \right){\bf 1}_A \right].
\end{align}
We note that this equality holds for any admissible $Y_T$ which is equal $P^*_T$ on $A^c$, such that $\xi Y$ is a martingale. Define  
$
Z :={\bf 1}_A  \sum_{i=1}^{n} U' \left(  P^{{*}}_{T_i} \right) \xi_{T_i}^{-1}  p_i.
$
Thus,
$
0\leq\EX \left[ \xi_T (P^{{*}}_T-Y_T)Z{\bf 1}_A \right],
$
which is equivalent to
\begin{equation}\label{eq:equiv}
0\leq \EX \left[ \xi_T (P^{{*}}_T-Y_T)Z|A \right].
\end{equation}
The theorem would follow if we can show that $Z$ is constant on $A$. Since
\begin{align*}
\EX\left[\xi_TP_T^{{*}}{\bf 1}_{A^c}\right]+\EX\left[\xi_TP^{{*}}_T{\bf 1}_A\right]&=\EX\left[\xi_TP_T^{{*}}\right]=\EX\left[\xi_TY_T\right]=\EX\left[\xi_TP_T^{{*}}{\bf 1}_{A^c}\right]+\EX\left[\xi_TY_T{\bf 1}_A\right]
\end{align*}
we have
$
\EX\left[\xi_T(P^{{*}}_T-Y_T)|A\right]= 0.
$
Hence, (\ref{eq:equiv}) implies
\begin{equation}
\label{geq0}
0\leq \EX\left[\xi_T(P_T-Y_T^{{*}})\tilde{Z}|A\right],
\end{equation}
with $\tilde{Z}:=(Z-\EX\left[Z|A\right]){\bf 1}_A$. By Lemma \ref{xdelta} this entails that 
\begin{equation} \label{eq:exp}
0\leq \EX\left[X\tilde{Z}|A\right]
\end{equation}
for any bounded $X$ with $\EX\left[X|A\right]=0$ satisfying $X\leq 0$ on $P^{{*}}_T=0$.

Now if $\tilde{Z}\geq 0$ or $\leq 0$ on $A$ we have that $\tilde{Z}=0$ on $A$ and we are done (since $\tilde{Z}=Z-\EX[Z]$ and therefore, $Z$ must be constant on $A$). On the other hand. if $\PP\left[\tilde{Z}<0|A\right]>0$ then
\begin{equation*}
\PP\left[P^{{*}}_T>0,\tilde{Z}<0|A\right]=\PP\left[\tilde{Z}<0|A\right]>0,
\end{equation*}
where the first equation holds as the wealth process $P^{{*}}_T$ is non-negative and $\PP\left[P^{{*}}_T=0|A\right]=0$ by the definition of $A$, since $U(y)$ is not differentiable at $y=0.$ For $a,b>0$ we can define 
$
X=\left\{-b{\bf 1}_{\tilde{Z}>0}+a{\bf 1}_{\tilde{Z}<0,P^{{*}}_T>0}\right\}{\bf 1}_A.
$
Then $X\leq 0$ on $P^{{*}}_T=0$. Choose $b,a> 0$ such that $\EX\left[X|A\right]=0$. Then by (\ref{eq:exp})
\begin{align*}
0&\leq \EX\left[X\tilde{Z}|A\right]=-b\EX\left[\tilde{Z}^+|A\right]-a\EX\left[\tilde{Z}^-{\bf 1}_{P^*_T>0}|A\right].
\end{align*}
Hence, ${\bf 1}_A\tilde{Z}^+=0$ and thus ${\bf 1}_A\tilde{Z}=0$, since $\EX[\tilde{Z}\vert A]=0.$
By the definition of $\tilde{Z}$ above this entails that $Z$ is constant. To obtain the representation of $Z$, we recall our definition $\nu_t = \xi_t^{-1} U'(P_t^{{*}})$ (for $t \in \{T_1, \dots, T_n\}$) from the very beginning of this proof. %
\endproof

\begin{lemma}
	\label{xdelta}
	$
	0\leq \EX\left[X\tilde{Z}|A\right]$
	for any $\F_T$-measurable 
	bounded $X$ with $\EX\left[X|A\right]=0$ satisfying $X\leq 0$ on $P^{{*}}_T=0$.
\end{lemma}
\begin{proof}
	Note that (\ref{geq0}) implies that
	$$\EX \left[ \xi_T \bar{X} \tilde{Z}|A \right] \geq   0,$$ for any $\F_T$-measurable $\bar{X} \leq P^{{*}}_T$ such that $\EX \left[ \xi_T \bar{X}|A \right] =0$ and $\xi_T \bar{X}$ is bounded. (Since this implies by Assumption 1 and the martingale representation theorem on $\F_T$ that there exists a corresponding admissible $Y$ with $\xi_T Y_T ={\bf 1}_A\xi_T(P^{*}_T-\bar{X})+{\bf 1}_{A^c} P^{*}_T $ such that $\xi Y$ is a martingale.) 
	In particular, 
	$\EX \left[ \tilde{X} \tilde{Z}|A \right]\geq   0$ for any bounded $\tilde{X} $ being $\mathcal{F}_T$-measurable such that $\EX \left[ \tilde{X} |A\right] = 0$ and $\tilde{X} \leq P^{{*}}_T \xi_T$. 
	By definition the wealth process, $P^{{*}}$, is non-negative.
	By multiplying $ \tilde{X}$ below with the positive constant $\frac{\delta}{K}$, we see in particular that we must have
	\begin{align}\label{eq:1}
	\EX \left[\tilde{X}\tilde{Z} |A\right] \geq  0,
	\end{align}
	for all bounded $\F_T$-measurable $\tilde{X}$ with $\EX[\tilde{X}|A] = 0$, such that  $\tilde{X} \leq 0$ on $P_T^{{*}}\xi_T\leq \delta$ for a $\delta>0$, and  $\tilde{X}$ bounded by $K$ else.
	Since $K$ was arbitrary, (\ref{eq:1}) holds actually for any bounded $\F_T$-measurable $\tilde{X}$ with $\EX\left[\tilde{X}|A\right]=0$ and $\tilde{X}\leq 0$ on $P_T\xi_T\leq \delta$ for some $\delta>0$. 
	Finally,
	suppose that an $\F_T$-measurable bounded $X$ satisfies $X\leq 0$ if $P^{{*}}_T=0$, and $\EX\left[X_T|A\right]=0$. Then define
	\begin{equation*}
	X^{\delta}=\bigg(X-\hat{\delta}(\delta)\bigg){\bf 1}_{\xi_TP^{{*}}_T>\delta,X>0}+{\bf 1}_{X\leq 0}X
	\end{equation*}
	with $0\leq\hat{\delta}(\delta)$ chosen such that $\EX\left[X^{\delta}\right]=0$. The existence of $\hat{\delta}(\delta)\in[0,\infty)$ if $\delta$ is small is guaranteed by the intermediate value theorem. If we exclude the case that $X=0$, then $\hat{\delta}(\delta)\downarrow 0$ as $\delta\downarrow 0$. Furthermore, $X^{\delta}\leq 0$ on $\xi_TP_T^{{*}}\leq\delta$ so that $X^{\delta}$ satisfies (\ref{eq:1}). Hence,
	$
	0\leq \EX\left[X^{\delta}\tilde{Z}|A\right]\overset{\delta\downarrow 0}{\rightarrow}\EX\left[X\tilde{Z}|A\right].
	$
\end{proof}

\section{Dynamic programming approach with random time horizon}\label{Se:DPP}
Concavification has been widely applied to solve non-concave optimization problems, see e.g. 
\cite{BichuchSturm, CarassusPham, Carpenter, Chen,Larsen,Reichlin,Rieger2012,ross2004,Stadje} in various settings where the time horizon is fixed and the market is complete.
The concavification argument is based on the fact that concavified hall $U_c$ strictly dominates the initial function $U$ only in a union of finite number of open intervals and $U_c$ is affine in this union. The key idea is that in order to gain more expected utility, it is possible to put all the expensive states to the left points of these intervals in the concavification region, keeping the budget constraint unchanged.

In this section, we show that the concavification technique may no longer be applicable in settings with a random time horizon. Furthermore, we derive a dynamic programming principle for such a non-concave optimization.

We will start with the following useful lemma, where with a slight abuse of notation we write $\tau$ instead of $\tau\wedge T$. 
\begin{lemma}
Let $\overset{\sim}{\tau}$ have the same distribution as $\tau$ conditioned on $\tau >t$ and independent of $W$. Then we have
\begin{align*}
&\mathbb{E}[U(x+\int_0^{\tau}\pi_s\sigma_sdW_s^\QQ)|\mathcal{G}_t]={\bf 1}_{\tau\leq t}U(x+\int_0^{\tau}\pi_s\sigma_sdW_s^\QQ)+\mathbb{E}[U(x+\int_0^{\overset{\sim}{\tau}}\pi_s\sigma_sdW_s^\QQ)|\mathcal{F}_t]{\bf 1}_{\tau>t}.
\end{align*}
\end{lemma}
\proof
Let Y be bounded and $\mathcal{G}_t$-measurable. By Jeulin (2006), Lemma 4.4 $Y(\omega)=I_{\tau>t}X_t(\omega)+I_{\tau\leq t}g_t(\omega,\tau)$ for some $\mathcal{F}_t$-measurable random variable $X_t$ and some family of $\mathcal{F}_t\otimes\mathcal{B}([0,T])$-measurable random variables $g_t(\cdot,u);t\geq u$. Let $\overset{\sim}{\tau}$ have the same distribution as $\tau$ conditioned on $\tau>t$ and independent of $W$. Then we have
\begin{align*}
\mathbb{E}[U(x+\int_0^{\tau}\pi_s\sigma_sdW_s^\QQ)Y]=&\mathbb{E}[I_{\tau\leq t}U(x+\int_0^{\tau}\pi_s\sigma_sdW_s^\QQ)Y]+\mathbb{E}[I_{\tau>t}U(x+\int_0^{\tau}\pi_s\sigma_sdW_s^\QQ)Y]\\
=&\mathbb{E}[I_{\tau\leq t}U(x+\int_0^{\tau}\pi_s\sigma_sdW_s^\QQ)Y]+\mathbb{E}[I_{\tau>t}U(x+\int_0^{\tau}\pi_s\sigma_sdW_s^\QQ)X_t]\\
=&\mathbb{E}[I_{\tau\leq t}U(x+\int_0^{\tau}\pi_s\sigma_sdW_s^\QQ)Y]+\mathbb{E[}\mathbb{E}[I_{\tau>t}U(x+\int_0^{\tau}\pi_s\sigma_sdW_s^\QQ)|\mathcal{F}_t]X_t]\\
=&\mathbb{E}[I_{\tau\leq t}U(x+\int_0^{\tau}\pi_s\sigma_sdW_s^\QQ)Y]+\mathbb{E}[\frac{I_{\tau>t}\mathbb{E}[I_{\tau>t}U(x+\int_0^{\tau}\pi_s\sigma_sdW_s^\QQ)|\mathcal{F}_t]}{P(\tau>t)}X_t]\\
=&\mathbb{E}[I_{\tau\leq t}U(x+\int_0^{\tau}\pi_s\sigma_sdW_s^\QQ)Y]+\mathbb{E}[I_{\tau>t}\mathbb{E}[U(x+\int_0^{\overset{\sim}{\tau}}\pi_s\sigma_sdW_s^\QQ)|\mathcal{F}_t]X_t]\\
=&\mathbb{E}[I_{\tau\leq t}U(x+\int_0^{\tau}\pi_s\sigma_sdW_s^\QQ)Y]+\mathbb{E}[I_{\tau>t}\mathbb{E}[U(x+\int_0^{\overset{\sim}{\tau}}\pi_s\sigma_sdW_s^\QQ)|\mathcal{F}_t]Y]\\
=&\mathbb{E}\bigg[\Big\{I_{\tau\leq t}U(x+\int_0^{\tau}\pi_s\sigma_sdW_s^\QQ)]+I_{\tau>t}\mathbb{E}[U(x+\int_0^{\overset{\sim}{\tau}}\pi_s\sigma_sdW_s^\QQ)|\mathcal{F}_t]\Big\}Y\bigg],
\end{align*}
from which the lemma follows by the definition of a conditional expectation.\endproof
Let $\overset{\sim}{\tau}$ have the same distribution as $\tau$ conditioned on $\tau >t$. Let us define
\begin{align*}
V(t,x)&=\underset{(\pi_s)_{t\leq s\leq T}}{\esssup}\mathbb{E}\bigg[U(P_{\tau\wedge T}^{t,\pi ,x}){\bf 1}_{\tau >t}|\mathcal{G}_t\bigg]=\underset{(\pi_s)_{t\leq s\leq T}}{\esssup}\mathbb{E}\bigg[U(x+\int_t^{\overset{\sim}{\tau}}\pi_s \sigma_s dW_s^\QQ)|\mathcal{F}_t\bigg]{\bf 1}_{\tau >t}
\end{align*}
and $\overset{\sim}{V}(t,x)=\underset{(\pi_s)_{t\leq s\leq T}}{\esssup}\mathbb{E}[U(P_{\tau\wedge T}^{\tau\wedge t,\pi,x})|\mathcal{G}_t]=U(x){\bf 1}_{\tau\leq t}+V(t,x){\bf 1}_{\tau >t}$. We want to find $\overset{\sim}{V}(0,x)$. Below we show that $\overset{\sim}{V}(t,x)$ follows the usual dynamic programming principle. 
\begin{proposition}[Dynamic Programming]\label{Pro:DP}
	For any $0\leq t\leq t'\leq T$, we have
\begin{equation*}
\overset{\sim}{V}(t,x)=\underset{(\pi_s)_{t\leq s\leq t'}}{\esssup}\mathbb{E}\bigg[\overset{\sim}{V}(t',P_{t'\wedge\tau}^{\tau\wedge t,\pi,x})|\mathcal{G}_t\bigg].
\end{equation*}
\end{proposition}
\begin{proof}
	It is
\begin{align*}
\overset{\sim}{V}(t,x)&=\underset{(\pi_s)_{t\leq s\leq t'}}{\esssup}\mathbb{E}\bigg[U(x){\bf 1}_{\tau\leq t}+U(x+\int^{\tau}_t\pi_s\sigma_s dW_s^\QQ){\bf 1}_{t<\tau\leq t'}\\
&\hspace{0.55cm}+\underset{(\pi_s)_{t'\leq s\leq T\wedge\tau}}{\esssup}\mathbb{E}\bigg[{\bf 1}_{\tau >t'}U(x+\int_{t}^{t'}\pi_s\sigma_s dW_s^\QQ+\int_{t'}^{T\wedge\tau}\pi_s\sigma_sdW_s^\QQ)|\mathcal{G}_{t'}\bigg]\bigg |\mathcal{G}_t\bigg]\\
&=\underset{(\pi_s)_{t\leq s\leq t'}}{\esssup}\mathbb{E}\bigg[\bigg\{U(P_{\tau\wedge T}^{\tau\wedge t,\pi ,x}){\bf 1}_{\tau\leq t'}+\underset{(\pi_s)_{t'\leq s\leq \tau\wedge T}}{\esssup}{\bf 1}_{\tau >t'}\mathbb{E}\bigg[U(P_{t'}^{\tau\wedge t,\pi ,x}+\int_{t'}^{T\wedge\tau}\pi_s \sigma_s dW_s^\QQ)|\mathcal{G}_{t'}\bigg]\bigg\}\bigg|\mathcal{G}_t\bigg]\\
&=\underset{(\pi_s)_{t\leq s\leq t'}}{\esssup}\mathbb{E}\bigg[U(P_{t'\wedge \tau}^{\tau\wedge t,\pi ,x}){\bf 1}_{\tau\leq t'}+V(t',P_{t'}^{\tau\wedge t,\pi ,x}){\bf 1}_{\tau >t'}\bigg|\mathcal{G}_t\bigg]=\underset{(\pi_s)_{t\leq s\leq t'}}{\esssup}\mathbb{E}\bigg[\overset{\sim}{V}(t',P_{t'\wedge\tau}^{\tau\wedge t,\pi ,x})\bigg |\mathcal{G}_t\bigg].
\end{align*}\end{proof}

We are now in a position to show that contrary to the case with a certain time horizon, concavifying the utility function is not applicable when the investment horizon is random. In other words, replacing $U$ with $U^c$ in the optimization \eqref{eq:general_problem} leads to a super-optimal strategy.
To this end, we need to investigate smoothness and concavity of the value function of the {\it certain} time horizon optimization problem
\begin{equation}
\bar{V}(t,x):=\sup_{\pi \in {\Pi}(t,x)} \EE[U(P_T)\vert P_t=x].
\label{eq:Vprob}
\end{equation}
Smoothness and concavity of the value function has been also studied in \cite{Biane2011} by working on the dual control problem and the dual HJB equation under the following assumption which we in the sequel will make as well \footnote{It is shown in \cite{Biane2019} that the continuity assumption of $U^c$ can be dropped under  a H\"older continuity condition (Theorem 4.2) by using the comparison principle of PDEs
for the dual control problem.}:

\vspace{2mm}
\noindent{\bf Assumption (H)}: $U(0)=U^c(0)=0$, $U^c(\infty)=\infty$ and $U^c$ is strictly increasing and 
$
  U^c(x) \le C (1+x^p)
$
for some constant $C>0$ and $0<p<1$.

\begin{proposition}\label{Pro:value} 
Under Assumption (H), the value function $\bar{V}(t,x)$ of Problem \eqref{eq:Vprob} is strictly concave and strictly increasing and $C^{1,2}$ in $[0,T)\times [0,\infty) $. Furthermore, $\bar{V}(T,x)=U^c(x)$, $\bar{V}(t,0)=0$ and $\bar{V}(t,x)\le \wt{C}(1+x^p)$ for some positive constant $\wt{C}$ and $\bar{V}(t,x)$ satisfies the Inada's condition at zero and infinity.
\end{proposition}
\proof By Theorem 4.1 in \cite{Reichlin} the concavification argument can be applied and $U$ can be replaced by its concave hull $U^c$. By assumption, $U^c$ is increasing and concave and it follows that $\bar{V}$ is strictly increasing, in $C^{1,2}$ and satifies the growth condition by applying Theorem 3.8 in  \cite{Biane2011}. An inspection of the proof of Theorem 3.8  together with Lemma 3.6 \cite{Biane2011} also confirms that $\bar{V}$ satisfies the Inada condition at $0$ and infinity. 
\endproof

\begin{proposition}\label{Pro:nonconcave} 
Assume that Assumption (H) holds and the concavification region $\{U<U^c\}$ contains an interval $(0,\eta)$ for some $\eta>0$\footnote{This assumption is satisfied in the option compensation problem with power utility, see Section \ref{Se:option}}. Define  
$\underset{\pi}{\sup}\,\mathbb{E}[U^c(P^{\pi,x}_{\tau\wedge T})]=A_0\quad \mbox{and}\quad \underset{\pi}{\sup}\,\mathbb{E}[U(P^{\pi,x}_{\tau\wedge T})]=B_0.$ Then $A_0>B_0$.
\end{proposition}
\proof 
By the dynamic programming principle it is sufficient to show that on a non-zero set
\begin{align*}
A_{T_{n-2}}&:=\underset{(\pi_s)_{T_{n-2}\leq s\leq T_{n}}}{\esssup}\mathbb{E}[U^c(P^{\pi}_{\tau\wedge T})|\mathcal{G}_{T_{n-2}}]>\underset{(\pi_s)_{T_{n-2}\leq s\leq T_{n}}}{\esssup}\mathbb{E}[U(P^{\pi}_{\tau\wedge T})|\mathcal{G}_{T_{n-2}}]=:B_{T_{n-2}}.
\end{align*}
Let us remark that in our complete market setting, the market price density $\xi$ is atomless and $U$ is continuous by assumption. In particular, by Proposition \ref{Pro:value} and the concavification techniques in the certain maturity case (see Section 5 of  \cite{Reichlin}), the last period value function is given by
\begin{align*}
V_{T_{n-1}}&=\underset{(\pi_s)_{T_{n-1}\leq s\leq T_{n}}}{\esssup}\mathbb{E}[U^c(P^{\pi}_{\tau\wedge T})|\mathcal{G}_{T_{n-1}}]=\underset{(\pi_s)_{T_{n-1}\leq s\leq T_{n}}}{\esssup}\mathbb{E}[U(P^{\pi}_{\tau\wedge T})|\mathcal{G}_{T_{n-1}}],
\end{align*}
which is strictly increasing and strictly concave and $V_{T_{n-1}}(0)=U(0)=0$. Therefore, 
\begin{equation*}
p_{n-1} U+p_{n}V_{T_{n-1}}\leq (p_{n-1} U+p_{n}V_{T_{n-1}})^c\leq p_{n-1}U^c+p_{n}V_{T_{n-1}}.
\end{equation*}
It can be observed that for an $\wt{\varepsilon}>0$
$$
(0,\wt{\varepsilon})\subset\{p_{n-1}U+p_{n}V_{T_{n-1}}<\big(p_{n-1}U+p_{n}V_{T_{n-1}}\big)^c\}\subset\{p_{n-1}U+p_{n}V_{T_{n-1}}<p_{n-1}U^c+p_{n}V_{T_{n-1}}\}=\{U<U^c\}.
$$
It then follows
\begin{align*}
A_{T_{n-2}}&=\sum_{i\le n-2} {\bf 1}_{{\tau} = T_i} U^c(P^{\pi}_{T_{i}})+ {\bf 1}_{\tau > T_{n-2}} \underset{(\pi_s)_{T_{n-2}\leq s\leq T_{n-1}}}{\esssup}\mathbb{E}[p_{n-1}U^c(P^{\pi}_{T_{n-1}})+p_{n}V_{T_{n-1}}(P^{\pi}_{T_{n-1}})|\mathcal{G}_{T_{n-2}}]/(p_{n-1}+p_{n})\\
&\geq\sum_{i\le n-2} {\bf 1}_{\tau = T_i} U^c(P^{\pi}_{T_{i}})+{\bf 1}_{\tau > T_{n-2}}  \mathbb{E}[p_{n-1}U^c(P^{\pi^*}_{T_{n-1}})+p_{n}V_{T_{n-1}}(P^{\pi^*}_{T_{n-1}})|\mathcal{G}_{T_{n-2}}]/(p_{n-1}+p_{n})\\
&>\sum_{i\le n-2} {\bf 1}_{\tau = T_i} U(P^{\pi}_{T_{i}})+{\bf 1}_{\tau > T_{n-2}}  \mathbb{E}[\big(p_{n-1}U(P^{\pi^*}_{T_{n-1}})+p_{n}V_{T_{n-1}}(P^{\pi^*}_{T_{n-1}})\big)^c|\mathcal{G}_{T_{n-2}}]/(p_{n-1}+p_{n})\\
&=\sum_{i\le n-2} {\bf 1}_{\tau = T_i} U(P^{\pi}_{T_{i}})+ {\bf 1}_{\tau > T_{n-2}} \underset{(\pi_s)_{T_{n-2}\leq s\leq T_{n-1}}}{\esssup}\mathbb{E}[p_{n-1}U(P^{\pi^*}_{T_{n-1}})+p_{n}V_{T_{n-1}}(P^{\pi^*}_{T_{n-1}})|\mathcal{G}_{T_{n-2}}]/(p_{n-1}+p_{n})\\
&=B_{T_{n-2}}
\end{align*}
with $\pi^*=\underset{(\pi_s)_{T_{n-2}\leq s\leq T_{n-1}}}{\mbox{argsup}}\mathbb{E}[p_{n-1}U(P^{\pi}_{T_{n-1}})+p_{n}V_{T_{n-1}}(P^{\pi}_{T_{n-1}})|\mathcal{G}_{T_{n-2}}]$. 
The strict inequality above holds because
$
\big(p_{n-1}U+p_{n}V_{T_{n-1}}\big)^c$ 
is not affine on the set $\{U<U^c\}$ and is strictly concave on an interval $(0,\varepsilon)\subset\{U<U^c\}$ for some $\varepsilon>0$ due to Inada's condition at zero of $V_{T_{n-1}}$ (see Lemma \ref{Le:conf}). Hence, by a Merton-Lagrange-type analysis, $P^{\pi^*}_{T_{n-1}}$ takes values with positive probability in a non-zero set where $
\big(p_{n-1}U+p_{n}V_{T_{n-1}}\big)^c<p_{n-1}U^c+p_{n}V_{T_{n-1}}$.
\endproof
It follows from Proposition \ref{Pro:nonconcave} that concavification techniques (as for instance in \cite{Reichlin}) cannot be directly applied to $U$ when the time horizon is random. The non-concave optimization in this case can however still be solved by a recursive procedure which is established by Proposition \ref{Pro:DP}. This will be explicitly illustrated in the next section.

\section{Example for power utility function} \label{ch:examples}
In this section, we illustrate our main results established in the previous sections. In particular, we consider for $0 \leq \tau \le T$ a discrete random variable, i.e., there are times $T_0 := 0 < T_1 < T_2 < \dots < T_n = T$ and probabilities $0 < p_i < 1$ for $1 \leq i \leq n$ with $\sum_{i=1}^{n} p_i = 1$ such that
$
\PP \left( \tau = T_i \right) = p_i,$ for $1 \leq i \leq n.$ 
For simplicity, we assume that $\theta$ and $r$ are  constant and we choose a power (CRRA) utility, i.e., 
\begin{align}
U(x) := \frac{x^{1-\gamma}}{1-\gamma}, \text{ for } \, 0< \gamma<1.\label{eq:powerU}
\end{align}

\subsection{Concave optimization with power utility} \label{ch:Merton_problem}

Note first that since $U$ is strictly concave we have that $I=(U')^{-1}$. {By Theorem \ref{thm:adapted_process}, we need to find an {adapted process} $\nu\ge 0$ with $\nu_0=U'(x)$ such that the process $\xi P^{x,\pi^{(n)}(x,I(\nu_{T_1} \xi_{T_1}),\cdots, I(\nu_{T_1}\xi_{T_n}))}$ generated by the $n+1$-tuple $(x,I(\nu_{T_1} \xi_{T_1}),\cdots, I(\nu_{T_1}\xi_{T_n}))$ is a martingale and $ \sum_{i=1}^{n} p_i \nu_{T_i}$ is a constant}. As shown below, for such a CRRA utility function we can find a $\nu$ which is deterministic, in particular, $\sum_{i=1}^{n} p_i \nu_{T_i}$ is a constant. 

\begin{proposition}\label{pro:concavepower1}
For power utility $U$ defined in \eqref{eq:powerU}, the optimal solution $P^*$ generated by the $n+1$-tuple $(x,I(\nu_{T_1} \xi_{T_1}),\cdots, I(\nu_{T_n}\xi_{T_n}))$, where
\begin{align*} 
\nu_{T_j} = \left( \frac{ x }{ f\left(\frac{\gamma-1}{\gamma},0,T_j \right) } \right)^{-\gamma},\quad  1\le j\le n.
\end{align*} 
and $f(q,t,T) := \exp \left(  - q\int_{t}^{T} (r_s + \frac{1}{2} \theta^2_s) ds + q^2 \int_{t}^{T} \frac{\theta^2_s}{2} ds \right) $.
Furthermore, the optimal investment strategy is the Merton strategy, i.e.  
the optimal fraction of wealth invested in the risky asset at time $t$ is given by $\frac{\mu_t - r_t}{\gamma \sigma_t^2},$ which is independent of the distribution of the stopping time.
\end{proposition}
\proof We try to find the optimal solution $P^*$ generated by the $n+1$-tuple $(x,I(\nu_{T_1} \xi_{T_1}),\cdots, I(\nu_{T_1}\xi_{T_n}))$, 
where $\nu_{T_j}$, $j=1,\cdots,n$ are deterministically chosen such that the martingale condition is fulfilled.  By Lemma \ref{lemma:Thais_lemma1}  
and the martingale condition we have for $1\le i\le j\le n$
\begin{align*}
\xi_{T_i}  I(\nu_{T_i} \xi_{T_i}) &= \EX \left[ \xi_{T_{j}}  I(\nu_{T_j} \xi_{T_j}) \big| \mathcal{F}_{ {T_i} } \right]=\xi_{T_i}  I(\xi_{T_i}) \nu_{T_j}^{-\frac{1}{\gamma}} f \left( \frac{\gamma-1}{\gamma}, T_i,T_j \right),  
\end{align*}
which yields 
\begin{align*} 
\nu_{T_i}^{-\frac{1}{\gamma}} = f\left(\frac{\gamma-1}{\gamma},{T_i},T_j \right)   \nu_{T_j}^{-\frac{1}{\gamma}},\quad 1\le i\le j\le n.
\end{align*}
Hence,
\begin{align*} 
\nu_{T_j} = \left( \frac{ x }{ f\left(\frac{\gamma-1}{\gamma},0,T_j \right) } \right)^{-\gamma},\quad  1\le j\le n.
\end{align*}
Using It\^{o}'s formula it can be checked directly that $\pi_t = \frac{\mu_t - r_t}{\gamma \sigma_t^2} P_t^*,$ 
which is the classical Merton strategy. \endproof

Hence, in the concave optimization problem the optimal portfolio selection is not affected by the presence of an uncertain time horizon, even though the value function is not identical to the one corresponding to the standard fixed-horizon case. This result can be considered as a confirmation of Merton \cite{Merton1971} and Richard \cite{Richard} and is aligned with the findings in \cite{Blanchet,BouchardPham}. 

\subsection{Non-concave optimization: recursive solution}\label{Se:option}


We consider the special choice of a non-concave objective function $U:\ree \to \ree\cup\{-\infty\}$ as in \eqref{eq:carpenter_utility}
, i.e., for given $K >0$ and $B > 0$:
\begin{align} \label{eq:carpenter_utility}
U(x) = \begin{cases}
u(K+\alpha (x-B)^{+}) & \text{ for } x \geq 0, \\
- \infty & \text{ else}. \end{cases}
\end{align}
where $u(x)=x^{1-\gamma}/(1-\gamma)$, with $0<\gamma<1$.
 We remark that although in almost all optimal control problems considered in the literature a fixed known time horizon is assumed, in reality a fixed maturity is typically not naturally given and the target date itself instead is often of random-type.
Hence, the problem considered in this chapter fits to all option type managerial compensation problems, for which in the case of a non-random time horizon there is already a rich literature in the finance \& OR literature going back to \cite{Carpenter,ross2004}.

Besides option type managearial compensation payoffs of the form (\ref{eq:carpenter_utility}) arise naturally for instance in flexibility rider insurance products which at the end of the life time of the policy holder, pay out a guarantee plus a particpation rate the latter depending on the returns in the stock market. In these products, the policy holder is allowed to influence the investment decision of the life insurance product. An example for such producs are in France for instance the life insurance products AXA Twin Star, in Germany the Swiss Life Champion and in the US, for example, Allianz Index Advanta, see \cite{Chen} and the references within. In this case $\alpha$, $K$, $B$ are the participation rate, the guarantee and the threshold for the participation, respectively.

By Proposition \ref{Pro:nonconcave}, a concavification procedure cannot be directly applied and we will solve the optimization by a recursive procedure. For comparison purpose, we introduce the concave envelope ${U}^c: \ree \to \ree\cup\{-\infty\}$ given by
\begin{align}
\begin{split} \label{eq:def_tildeu}
{U}^c(x) := \begin{cases}
- \infty & \text{ for } x < 0,\\
U(0) + U'(\hat{x}(B)) x &\text{ for } 0 \leq x \leq \hat{x}(B), \\
U(x) &\text{ for } x > \hat{x}(B),
\end{cases} 
\end{split}
\end{align}
where $\hat{x}(B) := \min \{ x > 0: U(x) = U^c(x)\}$. As in \cite{Stadje,Carpenter}, $\hat{x}(B)$ is defined by the following concavification equation: 
\begin{equation}
U(\hat{x}(B))-U(0)=U'(\hat{x}(B))\hat{x}(B).
\label{eq:conf}
\end{equation}

Note that $U^c$ dominates $U$ with equality for $x = 0$ and $ x \geq \hat{x}(B)$. 
Now, we are able to define the function $I: (0,\infty) \to [0,\infty)$ by
\begin{align} \label{eq:def_i}
I(y) := \left[\frac{1}{\alpha} \left(i \left( \frac{y}{\alpha} \right) - K \right) + B \right] {\bf 1}_{\{y < U'(\hat{x}(B))\}},
\end{align}
where $i(x)=x^{-1/\gamma}$ is the inverse of $u'$. 
We note that $I$ is the generalized inverse function of $({U}^c)^\prime$ in the sense that
$
y \in ({U}^c)^\prime(I(y)) \text{ for all } y > 0.
$

In our last period we already know $\tau$ so that the problem can be treated as a static non-concave EU maximization problem.

Given $P_{T_1}=x>0$, the wealth level at time $T_1$, the optimal terminal wealth of the conditional static problem is given by  
\begin{equation}
P_T^*=I(\lambda_T \xi_T)= \left[\frac{1}{\alpha} \left(i \left( \frac{\lambda_T \xi_T}{\alpha} \right) - K \right) + B \right] {\bf 1}_{\{\lambda_T \xi_T < U'(\hat{x}(B))\}}
\label{eq:XTcapr}
\end{equation}
where $\lambda_T$ is $\cF_{T_1}$-measurable and defined by the budget constraint $\EE[\xi_T\xi_{T_1}^{-1} P_T^*\vert \cF_{T_1}]=x$, see \cite{Carpenter,Stadje,ross2004}. 
The optimal wealth process is given by the following lemma: 
\begin{lemma}\label{Le:n2.0}
Given a realized wealth level at time $T_1$, the optimal wealth process on $(T_1,T]$ is given by ${P}_t^{*}=P_\zs{t,T}^*({\lambda_{T}\xi_{t}})$, where
\begin{align}
P_\zs{t,T}^*(y)&:=
f(1,t,T) \left(B-\frac{K}{{\alpha}}\right)\Phi[d(1,t,T,y)]\notag\\
&+\left(\frac{1}{{\alpha}}\right)^{1-\frac{1}{\gamma}}(y)^{-1/{\gamma}} f\left(1-\frac{1}{\gamma},{t},T\right)\Phi[d(1-1/\gamma,t,T,y )],
\label{eq:Indirect3}
\end{align}
where $\lambda_{T}$ satisfies the budget constraint at time $T_1$, $\Phi$ denotes the cumulative distribution function of the standard normal distribution and
\begin{equation}
d(q,t,T,\xi) =\frac{ \log \left(U'(\hat{x}(B))/\xi \right) +  \left( r + \frac{1}{2} \theta^2 \right) (T-t) }{ \theta \sqrt{T-t}} - q \theta{\sqrt{T-t}}.
\label{eq:d}
\end{equation}
\end{lemma}
\proof The lemma follows directly from \eqref{eq:XTcapr} and Lemma \ref{lemma:Thais_lemma2}. \endproof

{Note that the wealth process $P_\zs{t,T}^{*}(\lambda_T\xi_{t})$, expressed as a functional of the product $\lambda_T \xi_{t}$, depends on the wealth level at time $T_1$,  $P_{T_1}$, as the multiplier $\lambda_{T}=\lambda_{T}(P_{T_1})$ is characterized by the budget equation at time $T_1$. 


\begin{lemma}\label{Le:Indirect1}
The indirect value function ${V}_\zs{T_1}(x):=\EE[U(P_T^*)\vert P_{T_1}=x]$ is given by
\begin{align}
{V}_\zs{T_1}(x)=&\frac{1}{1-\gamma} f(q,{T_1},T) 
\left(\frac{1}{{\alpha}}\right)^{1-\frac{1}{\gamma}}({\lambda_{T}(x)\xi_{T_1}})^{1-1/{\gamma}} f\left(1-\frac{1}{\gamma},{T_1},T\right)\Phi[d(1-1/\gamma,T_1,T,\xi_{T_1}\lambda_{T}(x))]\notag\\
&+\frac{1}{1-\gamma}K^{1-\gamma}(1-\Phi[d(0,T_1,T,\xi_{T_1}\lambda_{T}(x))]).
\label{eq:VatT1}
\end{align}
\end{lemma}
\proof From \eqref{eq:XTcapr} we have 
$$
{V}_\zs{T_1}(x)=\EE[U(P_T^*)\vert P^*_{T_1}=x]=\EE\bigg[\frac{1}{1-\gamma}\bigg(\frac{\lambda_T(x)\xi_T}{\alpha}\bigg)^{(-1/\gamma)\times (1-\gamma)}{\bf 1}_{\{\lambda_T \xi_T < U'(\hat{x}(B))\}}\bigg]+\frac{K^{1-\gamma}}{1-\gamma} \EE\big[{\bf 1}_{\{\lambda_T \xi_T \ge U'(\hat{x}(B))\}}\big]
$$ 
and the explicit formula follows directly from Lemma \ref{lemma:Thais_lemma2}. \endproof
\begin{proposition}\label{Prop:Indirect}
${V}_\zs{T_1}(x)$ is a globally strictly concave function and its  first two derivatives are given by
\begin{equation}
{V}_\zs{T_1}'(x)=\lambda_{T}(x)\xi_\zs{T_\zs{1}}\quad \mbox{and} \quad V_\zs{T_1}^{''}(x)=\lambda_{T}'(x) \xi_\zs{T_\zs{1}}.
\label{eq:Indirect4}
\end{equation}
The inverse of marginal indirect value function $(V_\zs{T_1})'$ is given by
\begin{align}
I_\zs{T_1}(X)=&f(1,{T_1},T) \left(B-\frac{K}{{\alpha}}\right)\Phi[d(1,T_1,T,X)]\notag\\
&+\left(\frac{1}{{\alpha}}\right)^{1-\frac{1}{\gamma}}(X)^{-1/{\gamma}} f\left(1-\frac{1}{\gamma},{T_1},T\right)\Phi[d(1-1/\gamma,T_1,T,X)].\label{eq:Inverse}
\end{align}
\end{proposition}
%

\proof By differentiating the budget constraint
$$
x=f(1,T_1,T) \left(B-\frac{K}{{\alpha}}\right)\Phi[d(1,T_1,T,\lambda_{T}\xi_{T_1})]\notag\\
+\left(\frac{1}{{\alpha}}\right)^{1-\frac{1}{\gamma}}(\lambda_{T}\xi_{T_1})^{-1/{\gamma}} f\left(1-\frac{1}{\gamma},{T_1},T\right)\Phi[d(1-1/\gamma,T_1,T,\lambda_{T}\xi_{T_1})],
$$
we obtain $\frac{\d x}{\d \lambda_{T}}=A_1+A_2+A_3$, where
\begin{align*}
A_1&=\frac{-1}{\gamma} \left(\frac{1}{{\alpha}}\right)^{1-\frac{1}{\gamma}}\xi_{T_1}({\lambda_{T}\xi_{T_1}})^{-1/{\gamma}-1} f\left(1-\frac{1}{\gamma},{T_1},T\right)\Phi[d(1-1/\gamma,T_1,T,\lambda_{T}\xi_{T_1})],\\
A_2&=\left(\frac{1}{{\alpha}}\right)^{1-\frac{1}{\gamma}}({\lambda_{T}\xi_{T_1}})^{-1/{\gamma}} f\left(1-\frac{1}{\gamma},{T_1},T\right)\varphi[d(1-1/\gamma,T_1,T,\lambda_{T}\xi_{T_1})]\frac{-1}{\lambda_T \theta\sqrt{T-T_1}},\\
A_3&=f(1,{T_1},T) \left(B-\frac{K}{{\alpha}}\right)\varphi[d(1,T_1,T,\lambda_{T}\xi_{T_1})]\frac{-1}{\lambda_{T} \theta\sqrt{T-T_1}}.
\end{align*}
Similarly, by differentiating \eqref{eq:VatT1} we obtain that
\begin{align*}
\frac{\d x}{\d \lambda_{T}}{V}_\zs{T_1}'(x)=\lambda_{T}\xi_{T_1} A_1+ \frac{1}{1-\gamma}\lambda_{T}\xi_{T_1} A_2 -\frac{1}{1-\gamma}K^{1-\gamma}\varphi[d(0,T_1,T,\xi_{T_1}\lambda_{T})]\frac{-1}{\lambda_{T} \theta\sqrt{T-T_1}}.
\end{align*}

Note that from \eqref{eq:d} we have for any $q\in \bbr$,
\begin{align}
f(q,{T_1},T) \varphi[d(q,T_1,T,\lambda_{T}\xi_{T_1})]&=f(q,{T_1},T)\varphi[d(0,T_1,T,\lambda_{T}\xi_{T_1})-q \theta\sqrt{T-T_1}]\notag\\
&=f(q,{T_1},T)\varphi[d(0,T_1,T,\lambda_{T}\xi_{T_1})]e^{-\frac{1}{2}q ^2\theta^2({T-T_1})} e^{q d(0,T_1,T,\lambda_{T}\xi_{T_1})\theta\sqrt{T-T_1}}\notag\\
&=f(q,{T_1},T)\varphi[d(0,T_1,T,\lambda_{T}\xi_{T_1})] \frac{e^{q \log\bigg(\frac{U'(\hat{x}_B)}{\lambda_{T}\xi_{T_1}}\bigg)}}{f(q,{T_1},T)}\notag \\
&=\bigg(\frac{U'(\hat{x}_B)}{\lambda_{T}\xi_{T_1}}\bigg)^q\varphi[d(0,T_1,T,\lambda_{T}\xi_{T_1})].
\label{eq:keyrelation}
\end{align}
By direct calculation, we can represent ${V}_\zs{T_1}'(x)$ as 
\begin{align}
{V}_\zs{T_1}'(x)=&\frac{\d\lambda_{T} }{\d x} \bigg(\lambda_{T}\xi_{T_1} A_1+ \frac{1}{1-\gamma}\lambda_{T}\xi_{T_1} A_2 -\frac{1}{1-\gamma}K^{1-\gamma}\varphi[d(0,T_1,T,\xi_{T_1}\lambda_{T})]\frac{-1}{\lambda_{T} \theta\sqrt{T-T_1}}\bigg)\notag\\
=&
\frac{\d\lambda_{T} }{\d x} \lambda_{T}\xi_{T_1} (A_1+A_2+A_3)\notag\\
&+\frac{\d\lambda_{T} }{\d x}\bigg((\frac{1}{1-\gamma}-1)\lambda_{T}\xi_{T_1} A_2 -\frac{1}{1-\gamma}K^{1-\gamma}\varphi[d(0,T_1,T,\xi_{T_1}\lambda_{T})]\frac{-1}{\lambda_{T} \theta\sqrt{T-T_1}}-\lambda_{T}\xi_{T_1}A_3\bigg)\notag\\
=&
\lambda_{T}\xi_{T_1} +\frac{\d\lambda_{T} }{\d x}\bigg(\frac{\gamma}{1-\gamma}\lambda_{T}\xi_{T_1} A_2 -\frac{K^{1-\gamma}}{1-\gamma}\varphi[d(0,T_1,T,\xi_{T_1}\lambda_{T})]\frac{-1}{\lambda_{T} \theta\sqrt{T-T_1}}-\lambda_{T}\xi_{T_1}A_3\bigg).\label{eq:VVV}
\end{align}
By applying \eqref{eq:keyrelation} with $q=1$ and $q=1-\frac{1}{\gamma}$ we obtain
\begin{align*}
f(1,{T_1},T) \varphi[d(1,T_1,T,\lambda_{T}\xi_{T_1})]&=\bigg(\frac{U'(\hat{x}_B)}{\lambda_{T}\xi_{T_1}}\bigg)\varphi[d(0,T_1,T,\lambda_{T}\xi_{T_1})],\\
f(1-\frac{1}{\gamma},{T_1},T) \varphi[d(1-\frac{1}{\gamma},T_1,T,\lambda_{T}\xi_{T_1})]&=\bigg(\frac{U'(\hat{x}_B)}{\lambda_{T}\xi_{T_1}}\bigg)^{1-\frac{1}{\gamma}}\varphi[d(0,T_1,T,\lambda_{T}\xi_{T_1})].
\end{align*}
It follows that
\begin{align}
 ({\lambda_{T}\xi_{T_1}})A_2&=
({\lambda_{T}\xi_{T_1}})\left(\frac{1}{{\alpha}}\right)^{1-\frac{1}{\gamma}}({\lambda_{T}\xi_{T_1}})^{-1/{\gamma}} \bigg(\frac{U'(\hat{x}_B)}{\lambda_{T}\xi_{T_1}}\bigg)^{1-\frac{1}{\gamma}}\varphi[d(0,T_1,T,\lambda_{T}\xi_{T_1})]\frac{-1}{\lambda_T \theta\sqrt{T-T_1}}\notag\\
&
=
\bigg(\frac{U'(\hat{x}_B)}{\alpha}\bigg)^{1-\frac{1}{\gamma}}\varphi[d(0,T_1,T,\lambda_{T}\xi_{T_1})]\frac{-1}{\lambda_T \theta\sqrt{T-T_1}}\label{eq:A222}
\end{align}
and 
\begin{align}
({\lambda_{T}\xi_{T_1}})A_3&=
({\lambda_{T}\xi_{T_1}})\left(B-\frac{K}{{\alpha}}\right)\bigg(\frac{U'(\hat{x}_B)}{\lambda_{T}\xi_{T_1}}\bigg)\varphi[d(0,T_1,T,\lambda_{T}\xi_{T_1})]
\frac{-1}{\lambda_{T} \theta\sqrt{T-T_1}}\notag\\
&=\left(B-\frac{K}{{\alpha}}\right)U'(\hat{x}_B)\varphi[d(0,T_1,T,\lambda_{T}\xi_{T_1})]
\frac{-1}{\lambda_{T} \theta\sqrt{T-T_1}}.\label{eq:A222}
\end{align}
The bracket in \eqref{eq:VVV} can be expressed as 
\begin{align*}
\bigg(\underbrace{\frac{\gamma}{1-\gamma} \left(\frac{1}{{\alpha}}\right)^{1-\frac{1}{\gamma}}(U'(\hat{x}_B))^{1-1/\gamma}-\frac{1}{1-\gamma}K^{1-\gamma}
-(B-\frac{K}{\alpha}){U'(\hat{x}_B)})}_{A_4}\bigg)  \frac{-\varphi[d(0,T_1,T,\lambda_{T}\xi_{T_1})]}{\lambda_{T} \theta\sqrt{T-T_1}}.
\end{align*}
From \eqref{eq:carpenter_utility} we have $$U(0)=\frac{1}{1-\gamma}K^{1-\gamma},\quad \frac{\gamma}{1-\gamma}(\frac{1}{{\alpha}})^{1-\frac{1}{\gamma}}(U'(\hat{x}_B))^{1-1/\gamma}=\gamma U(\hat{x}_B),$$
and
\begin{align*}
(1-\gamma )U(\hat{x}_B)-\hat{x}_B U'(\hat{x}_B)&=(1-\gamma)\frac{(\alpha \hat{x}_B-\alpha B+K)^{1-\gamma} }{1-\gamma}-\hat{x}_B \alpha (\alpha \hat{x}_B-\alpha B+K)^{-\gamma} \\
&=
(B-\frac{K}{\alpha}){U'(\hat{x}_B)}.
\end{align*}
This implies that $A_4=U(\hat{x}_B)-U(0)-\hat{x}_B U'(\hat{x}_B)=0$ due to the concavification equation \eqref{eq:conf}. Hence,
$
{V}_\zs{T_1}'(x)={\lambda_{T}\xi_{T_1}}.
$
The above derivation also shows that  \eqref{eq:Inverse} defines the inverse of ${V}_\zs{T_1}.$\endproof

For a power utility function, it is straightforward to compute the optimal investment strategy in the period $[T_1,T)$ given the wealth level at time $T_1$, see e.g. \cite{Carpenter,Stadje}. Having determined the indirect utility function at time $T_1$, we now represent the optimization problem as
\begin{align} 
 \sup_{ (\pi_t)_\zs{t\in[0,T_1]}} \EE\bigg[pU(P_{T_1})+ (1-p) {V}_\zs{T_1}(P_{T_1})\bigg].
\label{eq:T1optdivi}
\end{align} 
Note that \eqref{eq:T1optdivi} is expressed as a non-concave optimization problem in a complete market. To solve it we look at its static version
\begin{align} 
 \sup_{P\in\cF_\zs{T_1} }\EE\bigg[pU(P)+ (1-p) {V}_\zs{T_1}(P)\bigg]=\sup_{P\in\cF_\zs{T_1} }\EE\bigg[U_1(P) {\bf 1}_{P\le B}+ U_2(P) {\bf 1}_{P> B}\bigg],
\label{eq:T1optdivi2}
\end{align} 
subject to the usual budget constraint $\EE[\xi_{T_1}P]\le x$, where $U_i, i=1,2$ are concave functions defined by
\begin{equation}
U_1(x):=pU(0)+ (1-p) {V}_\zs{T_1}(x),\quad U_2(x):=pU(x)+ (1-p) {V}_\zs{T_1}(x).
\label{eq:U12}
\end{equation}

Since in the last period the problem becomes static, the solution of the non-concave optimization \eqref{eq:T1optdivi2} is given by maximizing the concavified target function. Let $I_i,i=1,2$ be the corresponding inverse marginal utilities. The optimal wealth at $T_1$ is given by the following expression.  

\begin{proposition}\label{pro:XT1}
The optimal portfolio of Problem \eqref{eq:T1optdivi2} is given by
\begin{equation*}
P^*_\zs{T_1}={I}_\zs{2}({\lambda}\xi_\zs{T_1}){\bf 1}_\zs{\xi_\zs{T_1}<\wh{\xi}}+{I}_\zs{1}({\lambda}\xi_\zs{T_1}){\bf 1}_\zs{\xi_\zs{T_1}\ge\wh{\xi}}\,,
\end{equation*}
where $\wh{\xi}$ is defined by
\begin{equation}
{U}_\zs{2}({I}_\zs{2}({\lambda}\wh{\xi}))-
{U}_\zs{1}({I}_\zs{1}({\lambda}\wh{\xi}))
={\lambda}\wh{\xi} \bigg({I}_\zs{2}({\lambda}\wh{\xi})-{I}_\zs{1}({\lambda}\wh{\xi})\bigg),
\label{eq:commonline}
\end{equation}
and ${\lambda}$ is determined such that the budget constraint $\EE[\xi_\zs{T_1}P^*_\zs{T_1}]=x$ is satisfied.
\end{proposition}

Before presenting the proof let us remark that \eqref{eq:commonline} defines the linear line that is jointly tangent to the curves of ${U}_\zs{1}$ and ${U}_\zs{2}$.

\proof
For $\wh{\lambda}>0$ and $\xi>0$, consider the following Lagrangian 
 \begin{equation*}
\Psi(x):={U}_\zs{1}  	(x){\bf 1}_\zs{x< B}+{U}_\zs{2}  	(x){\bf 1}_\zs{x\ge B}-{\lambda} \xi x.
\end{equation*}
Note first that $\Psi$ is continuous and ${U}_\zs{i}$ attains maximum at ${I}_\zs{i}({{\lambda}}{\xi})$, $i=1,2$. Furthermore, it follows from \eqref{eq:U12} that ${I}_\zs{2}({{\lambda}}{\xi})>{I}_\zs{1}({{\lambda}}{\xi})$ for all ${\lambda}>0$ and $\xi>0$. Let $\xi_{B,1}:=\frac{U'_\zs{1}(B) }{{\lambda}}$. If $\xi\le \xi_{B,1}$, then ${I}_\zs{1}({{\lambda}}{\xi})>B$. Hence $\Psi$ is increasing in $[0,{I}_\zs{2}({{\lambda}}{\xi}))$ and decreasing in $[{I}_\zs{2}({{\lambda}}{\xi}),\infty)$. So ${I}_\zs{2}({{\lambda}}{\xi})$ is the maximizer when $\xi<\xi_{B,1}$. Similarly, for $\xi\ge \xi_{B,2}:=\frac{U'_\zs{2}(B) }{{\lambda}}>\xi_{B,1}$ we observe that
$\Psi$ is increasing in $[0,{I}_\zs{1}({{\lambda}}{\xi}))$ and decreasing in $[{I}_\zs{1}({{\lambda}}{\xi}),\infty)$. So ${I}_\zs{1}({{\lambda}}{\xi})$ is the maximizer for $\xi\ge \xi_{B,2}$. It remains to consider the case $\xi_{B,1}\le \xi\le \xi_{B,2}$. The global optimality of $\Psi$ results from the comparison of $\Psi({I}_\zs{2}({{\lambda}}{\xi}))$ and $\Psi({I}_\zs{1}({{\lambda}}{\xi}))$. To this end, consider the continuous function 
$$
f(\xi):=\Psi({I}_\zs{2}({{\lambda}}{\xi}))-\Psi({I}_\zs{1}({{\lambda}}{\xi}))=U_2({I}_\zs{2}({{\lambda}}{\xi}))-U_1({I}_\zs{1}({{\lambda}}{\xi}))-\lambda \xi({I}_\zs{2}({{\lambda}}{\xi})-{I}_\zs{1}({{\lambda}}{\xi})). 
$$
Obviously $f'(\xi)=-{\lambda}({I}_\zs{2}({{\lambda}}{\xi})-{I}_\zs{1}({{\lambda}}{\xi}))<0$, which implies that $f$ is decreasing in $\xi\in(0,\infty)$. Furthermore, noting that ${U}_\zs{1}(B)={U}_\zs{2}(B)$ we obtain
$$
f(\xi_{B,1})=
{U}_\zs{2}({I}_\zs{2}({{\lambda}}\xi_{B,1}))
-{U}_\zs{2}(B)-{U}'_\zs{2}({I}_\zs{2}({{\lambda}}{\xi_{B,1}}))({I}_\zs{2}({{\lambda}}{\xi_{B,1}})-B)>0,
$$
and 
$$
f(\xi_{B,2})=
{U}_\zs{1}(B)
-{U}_\zs{1}({I}_\zs{1}({{\lambda}}{\xi_{B,2}}))-{U}'_\zs{1}({I}_\zs{1}({{\lambda}}{\xi_{B,2}}))(B-{I}_\zs{1}({{\lambda}}{\xi_{B,1}})<0,
$$
because $U_1$ and ${U}_\zs{2}$ are strictly concave. Therefore, there exists $\wh{\xi}\in[\xi_{B,1},\xi_{B,2}]$ such that $f(\wh{\xi})=0$ which gives the concavification equation \eqref{eq:commonline}. Note that $f$ is strictly positive in $[\xi_{B,1},\wh{\xi})$ and strictly negative in $(\wh{\xi},\xi_{B,2}]$. The global maximizer of $\Psi$ is then given by ${I}_\zs{2}({{\lambda}}{\xi})$ if $\xi<\wh{\xi}$ or by ${I}_\zs{1}({{\lambda}}{\xi})$ if $\xi\ge\wh{\xi}$. The existence of $\lambda$ is not difficult to see.\endproof
\vspace{-2mm}
\begin{figure}[h]%
\centering\includegraphics[width=0.7\linewidth,height=7cm]{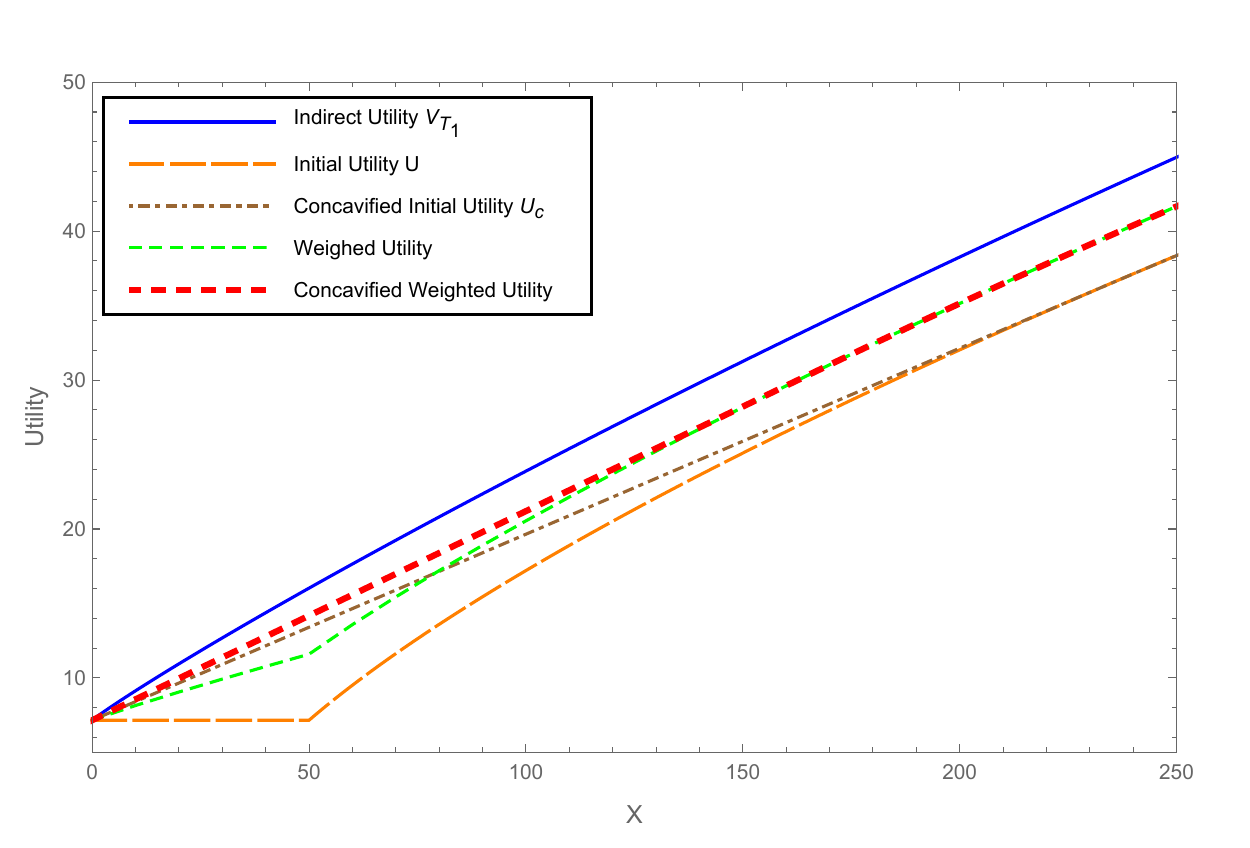}%
\caption{\small Weighted utility at time $T_1$ with $p=1/2$.}%
\label{Fig2}%
\end{figure}
\vspace{-2mm}
\subsection{Numerical illustration} \label{ch:numerical_illustration} 
We consider a classical Black-Scholes market with a risky asset $S$ and a bond $B$ 
and the $0 < T_1=5 < T=10$ such that
$
\PP \left( \tau = T_1 \right) = \PP \left( \tau = T  \right) = 1/2.
$
We assume that $\mu=0.08$ $r=0.03$, $\sigma=0.2$, $x_0=100$, $K=10$, $B=50$, $\gamma=0.3$, $\alpha=0.5$. We carry out a recursive procedure to determine the optimal solution for the non-concave problem with random time horizon $T\wedge \tau$.
\begin{center}
	\begin{figure}[h]
		\begin{subfigure}[b]{0.49\textwidth}
			\includegraphics[width=0.99\linewidth,height=6cm]{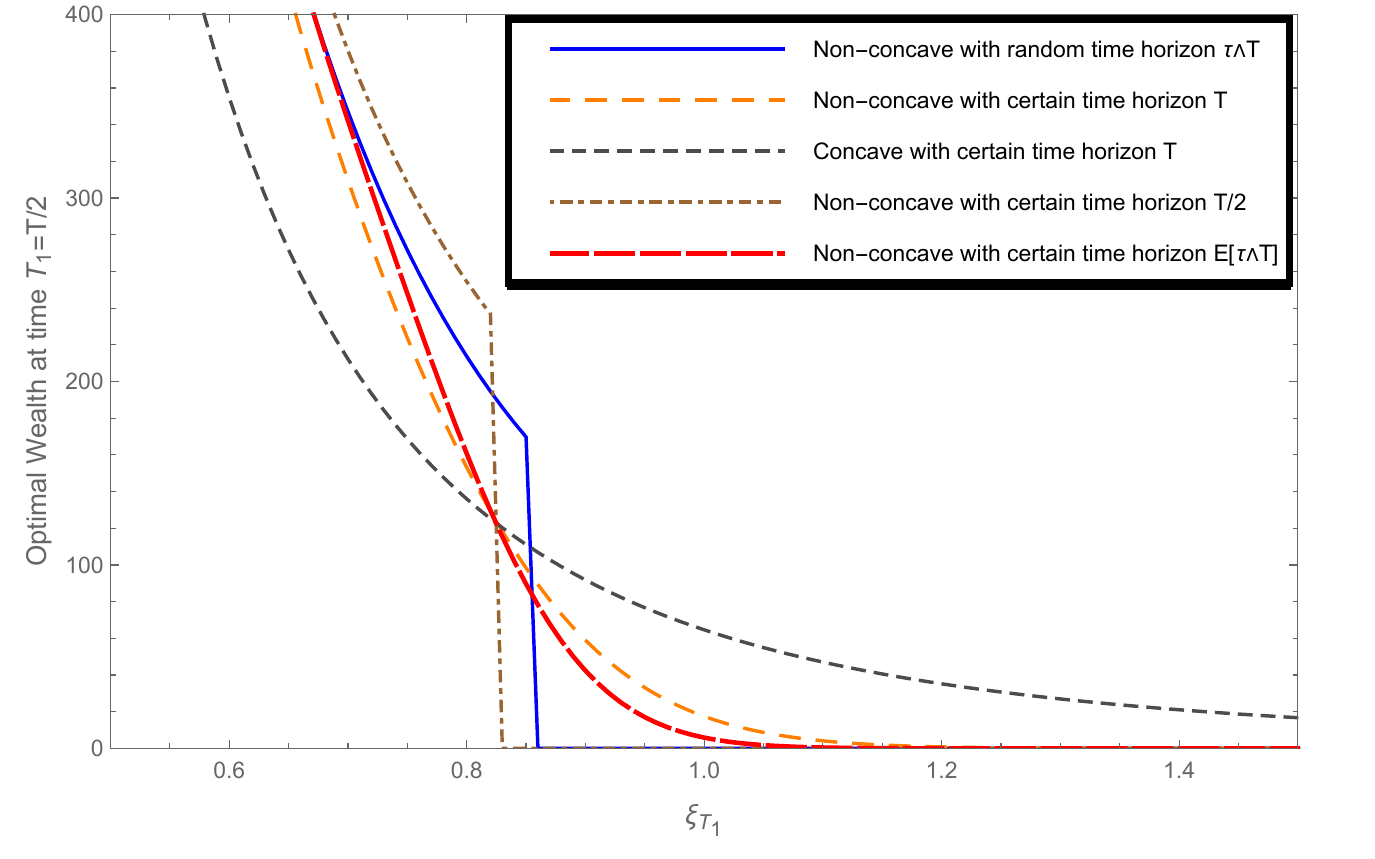}%
			\caption{$p=0.5$}%
	\label{Fig4}	
	\end{subfigure}%
		\begin{subfigure}[b]{0.49\textwidth}
			\includegraphics[width=0.99\linewidth,height=6cm]{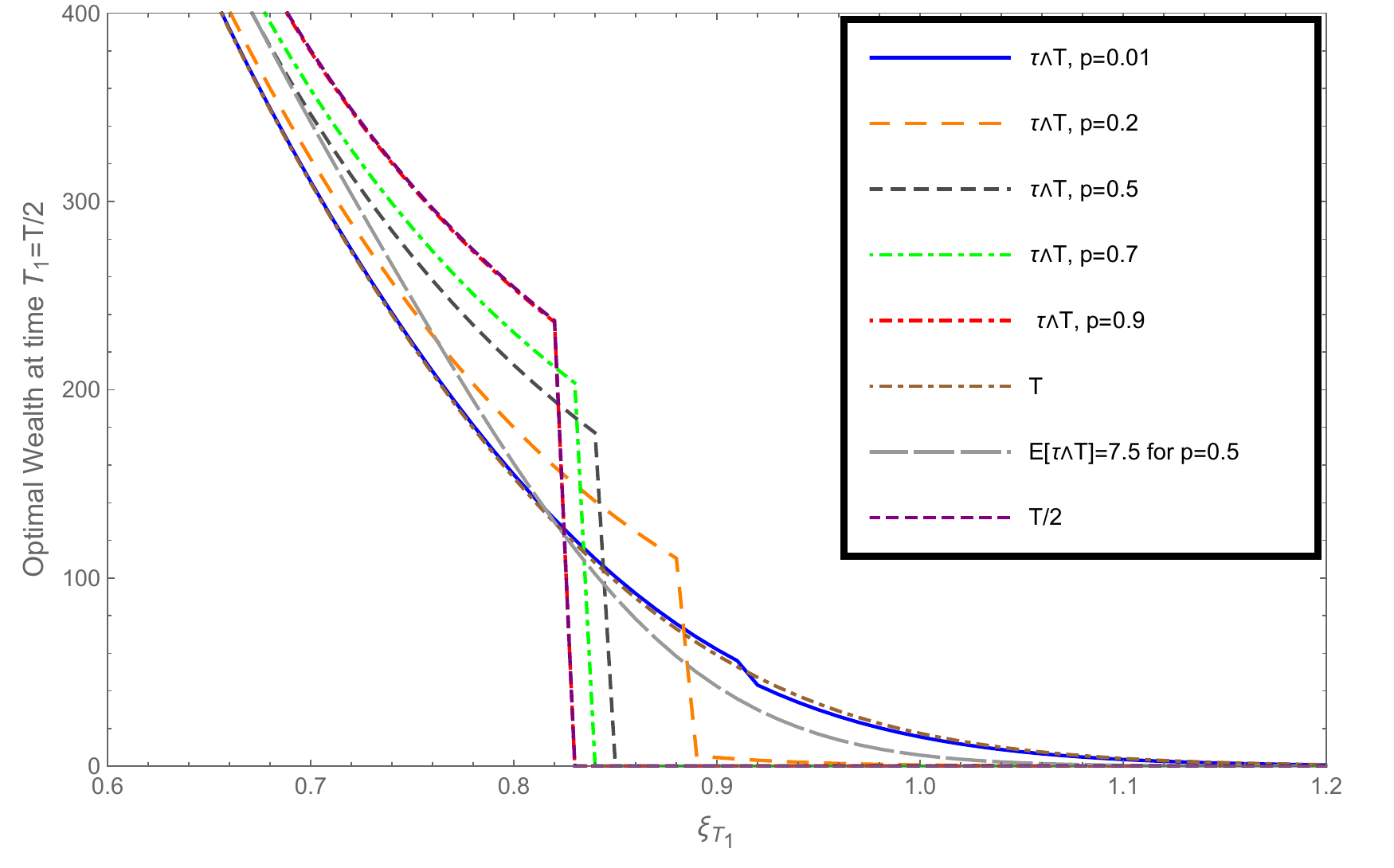}%
			\caption{Impact of $p$}
				\label{Fig41}
		\end{subfigure}%
		\caption{Optimal wealth at $T_1=T/2$}
				\label{Fig401}
	\end{figure}
\end{center}
Our numerical illustration relies on a Monte-Carlo simulation with 50000 paths of the market price density $\xi_{T_1}$ to determine the optimal multiplier $\lambda$ in the first period. This recursive procedure is computationally rather challenging. First, although the indirect value function of the last period can be computed in closed form in \eqref{eq:VatT1}, it implicitly depends on the price density $\xi_{T_1}$. Second, computation of the marginal utility functions ${I}_\zs{1}$, ${I}_\zs{2}$ of the corresponding concavified utilities is computationally intensive as concavification requires a root search step for each value of the market price density $\xi_{T_1}$. This is done using Brent's method with a careful choice of the starting values. 
Below, we numerically test and confirm the theoretical result established in Section \ref{Se:option}


In order to test the concavity of the weighted utility at time $T_1$, we plot in Figure \ref{Fig2} the indirect valued function at time $T_1$ defined in \eqref{eq:VatT1}. The graph numerically confirms the result in Proposition \ref{Prop:Indirect} that $V_{T_1}$ is strictly concave and dominates the initial utility $U$. In addition, the weighted utility defined by \eqref{eq:T1optdivi} is indeed non-concave and its concave hull is dominated by the indirect value function $V_{T_1}$. This implies that having a premature stopping time before $T$ leads to lower expected utility than the solution with certain time horizon $T$. In other words, this numerical example also confirms the result in Proposition \ref{Pro:nonconcave} that 
optimizing the concavified version of the utility function will lead to sub-optimality. 
%

 \begin{table}[h]%
\begin{center}%
 \begin{tabular}{l|ll}
 $p$ & $\lambda$ & $p\nu_{T_1}+(1-p)\nu_{T_2}$ \\
\hline
 0.1 & 0.17661847317422547 & 0.1766184731742255 \\
 0.2 & 0.17365388191218573 &  0.1736538819121857 \\
 0.3 & 0.170919 & 0.170919 \\
 0.4 & 0.16871475448916323 & 0.16871475448916323 \\
 0.5 & 0.165183 & 0.165183 \\
 0.6 & 0.1627175838160374 & 0.1627175838160374\\
 0.7 & 0.1598441738970805 & 0.1598441738970805\\
 0.8 & 0.15719801306953618 &  0.15719801306953618 \\
 0.9 & 0.15428390949982979 & 0.15428390949982979 \\
 \end{tabular}
 \caption{Weighted multiplier on the set $A$.}
 \label{Tab:A}
 \end{center}\vspace*{-2mm}
\end{table}
The optimal wealth at time $T_1$ is plotted in Figure \ref{Fig4} which exhibits an intermediate investment behavior between the non-concave problems with certain time horizon $T=\max\{T\wedge\tau(\omega)\}$ and $T/2=\min \{T\wedge \tau(\omega)\}$. In particular, it is higher (resp. lower) in good market scenarios but lower (resp. higher) in bad market states than the corresponding wealth of the certain time horizon problem $T$ (resp. $T/2$). In addition, there are ranges of intermediate market states in which the uncertain time wealth can be higher and lower than that of the non-concave problem with (certain) average time horizon $\EE[\tau\wedge T]=pT_1+(1-p)T=7.5$. As confirmed in Figure \ref{Fig41}, the larger (resp. smaller) the probability of exiting at the smallest time horizon value $T/2$, the riskier (resp. less risky) the investment behavior at time $T/2$. Furthermore, the random horizon problem converges to the extreme cases with certain horizon $T$ and $T/2$ when $p$ approaches to 0 and 1 respectively.

To further understand this effect, we plot in Figure \ref{Fig.51} the estimated density of the optimal wealth at time $T_1$ from 5000 simulations of the market price density $\xi_{T_1}$. It is interesting to observe that the distribution of the wealth at time $T_1$ of the non-concave optimization problems is right-skewed with a long right tail, indicating that the investor expects frequent small losses and a few large gains from the investment. A positively skewed distribution of investment returns is generally desirable by the agent with option-liked compensation payoff. In addition, the premature (before time $T$) exiting risk forces the investor to follow a portfolio that is of right-skewed and bimodal distribution with peaks of different heights. The bimodal structure can be explained by the concavification procedure at $T_1$, whereas the binomial distribution of the exiting time $\tau$ has significant impact on the amplitude between the two modes. The higher the probability $p$, the larger the amplitude. While the (certain) average time horizon portfolio is right-skewed and unimodal, the random time horizon portfolio, due to the option-liked compensation payoff at time $T_1$, is bimodal distributed which provides the investor flexibility of switching between the two local maximizers ${I}_\zs{1}$ and ${I}_\zs{2}$, depending on the market performance. If the concavified utility at time $T_1$ is affine in many open intervals, the corresponding wealth is intuitively expected to be of multimodal distribution. Again, when $p$ approaches to $0$ or $1$, the wealth distribution of the random horizon problem converges to the extreme cases with certain horizon $T$ and $T/2$.

\begin{center}
\begin{figure}[h]
      \begin{subfigure}[b]{0.45\textwidth}
\centering\includegraphics[width=0.9\linewidth,height=6cm]{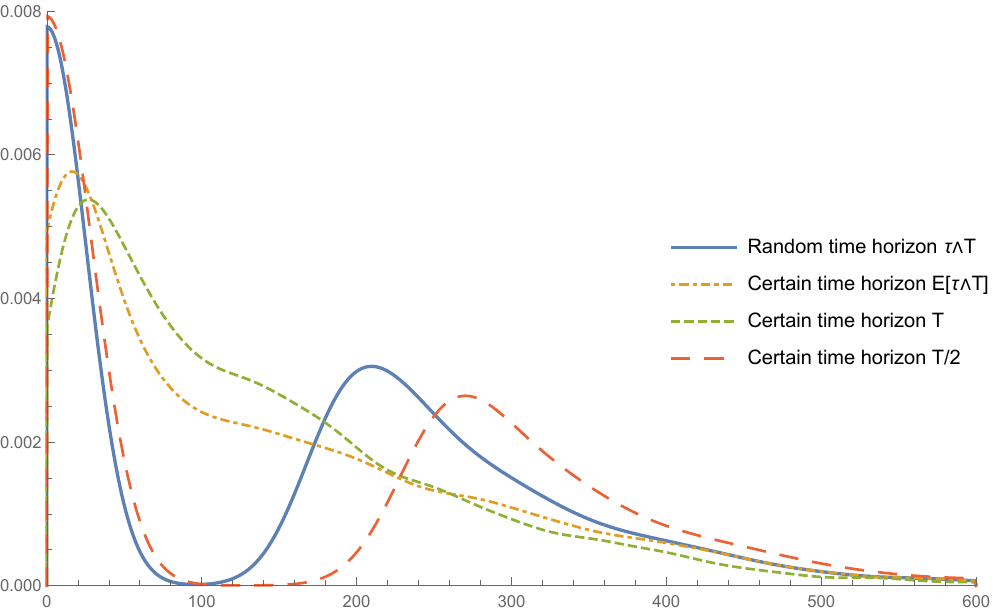}%
\caption{\small Effect of randomization with $p=0.5$.}%
	    \end{subfigure}%
   \begin{subfigure}[b]{0.45\textwidth}
     \centering	\includegraphics[width=0.9\linewidth,height=6cm]{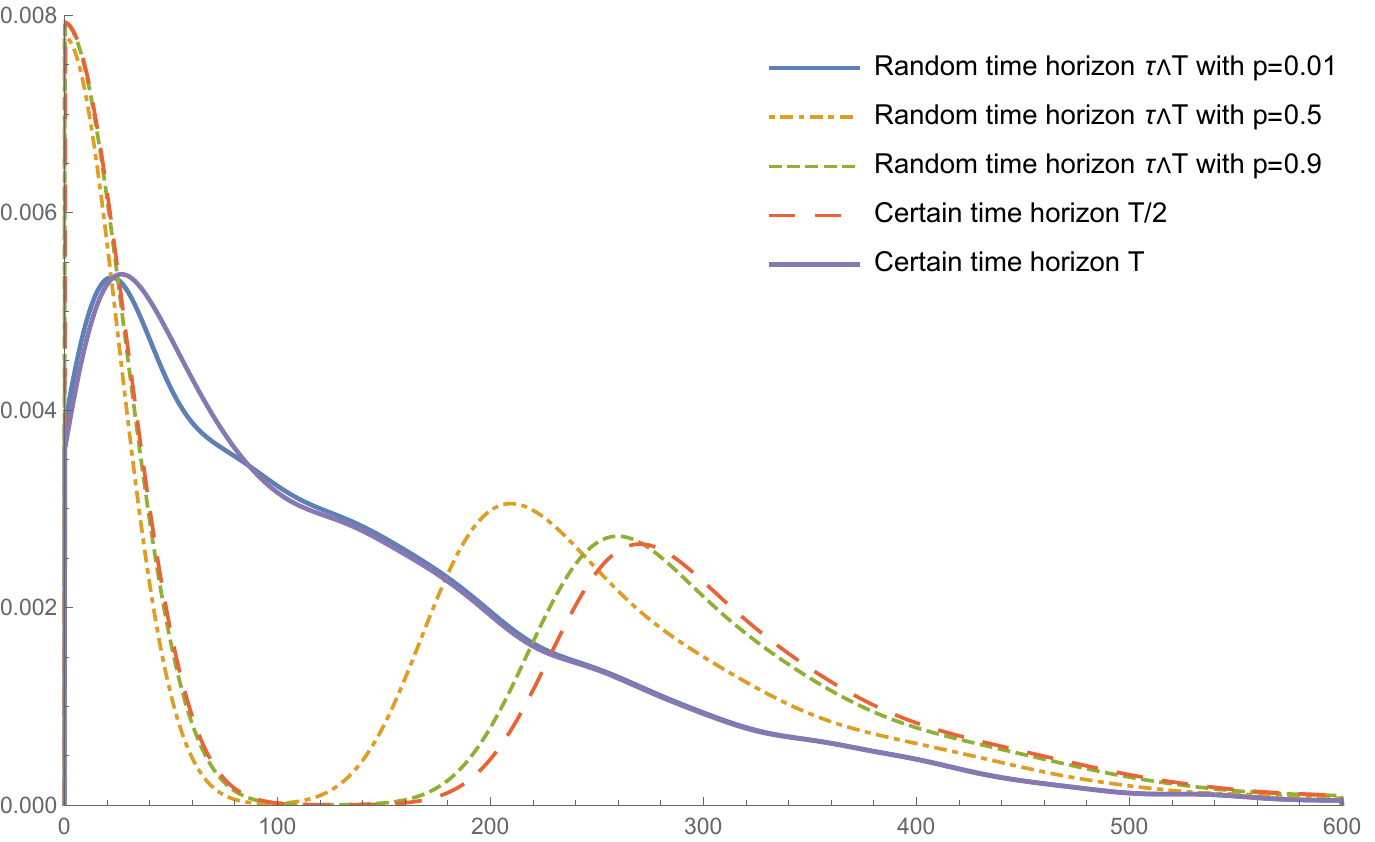}
		\caption{\small Effect of $p$. }
       \end{subfigure}%
      \caption{\small Estimated density of the optimal wealth at time $T_1=T/2=5$.}
	\label{Fig.51}
		 \end{figure}
	\end{center}
We now turn our attention to the impact of the time horizon uncertainty on the total expected utility. We first remark that in certain time horizon settings, it can numerically be shown that the value function of the concave and the non-concave problems is a convex function in the time horizon variable. Figure \ref{Fig.6} reports the impact of exiting probability $p=\PP(\tau =T/2)$ on the expected utility of the random time horizon $\tau \wedge T $ and the certain time horizon $\EE[\tau \wedge T]$. As shown in the right panel, the expected utility of the random horizon problem is always higher than that of the certain horizon problem, which is due to the convexity in time horizon of the value function and the fact that investment strategies for both cases with certain and uncertain horizon time horizon are identical and given by the Merton fraction (see Proposition \ref{pro:concavepower1}). 

The left panel of Figure \ref{Fig.6} reports the expected utility of the non-concave optimization setting. We observe a similar expected utility dominance of the uncertain time horizon problem over the certain time horizon problem when $p$ is close to $0$ and $1$. However, this effect is hard to see for intermediate values of $p$ for the given parameters. Unlike concave problems, the optimal investment strategy of the non-concave optimization problem significantly depends on the time horizon. 
\begin{center}
\begin{figure}
      \begin{subfigure}[b]{0.49\textwidth}
		\centering\includegraphics[width=0.9\columnwidth,height=6cm]{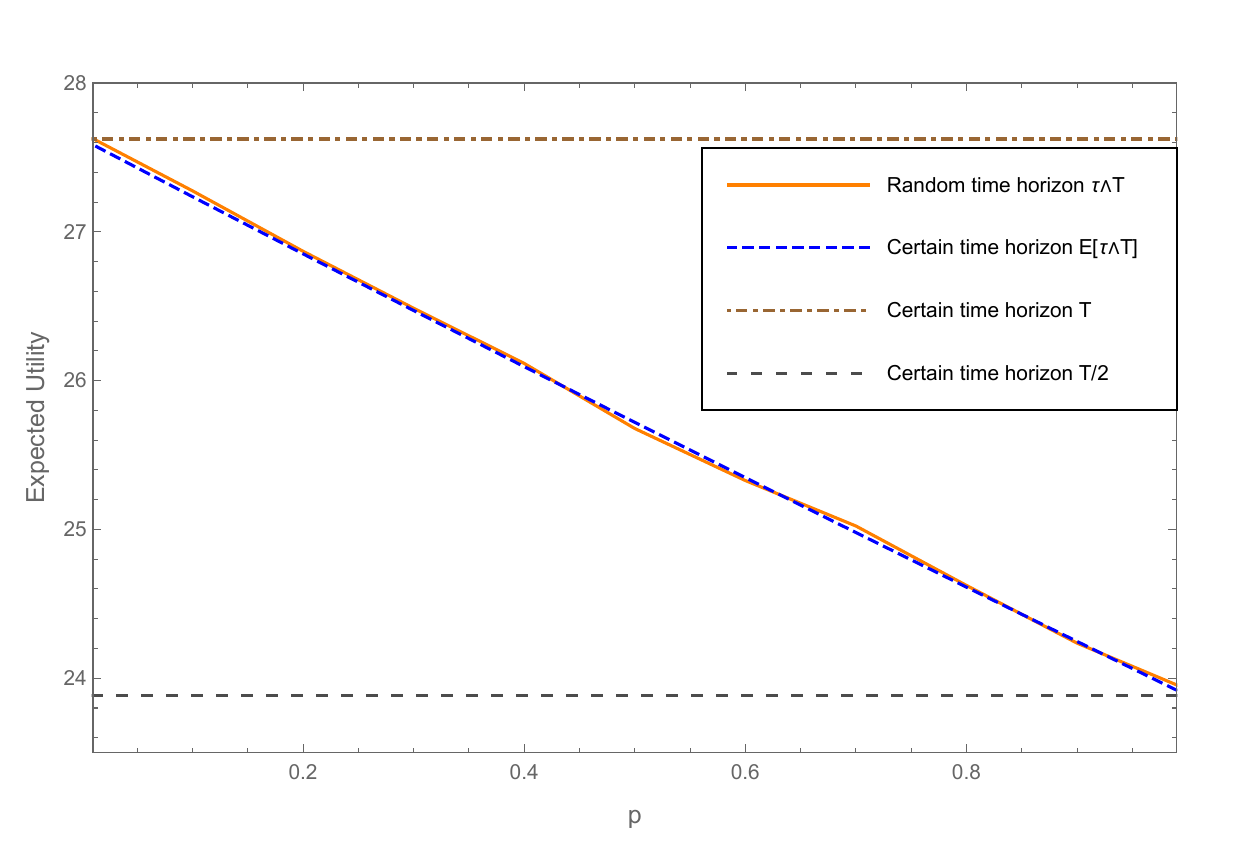}%
			\caption{\small Non-concave problem}
    \end{subfigure}%
   \begin{subfigure}[b]{0.49\textwidth}
     	\centering\includegraphics[width=0.9\columnwidth,height=6cm]{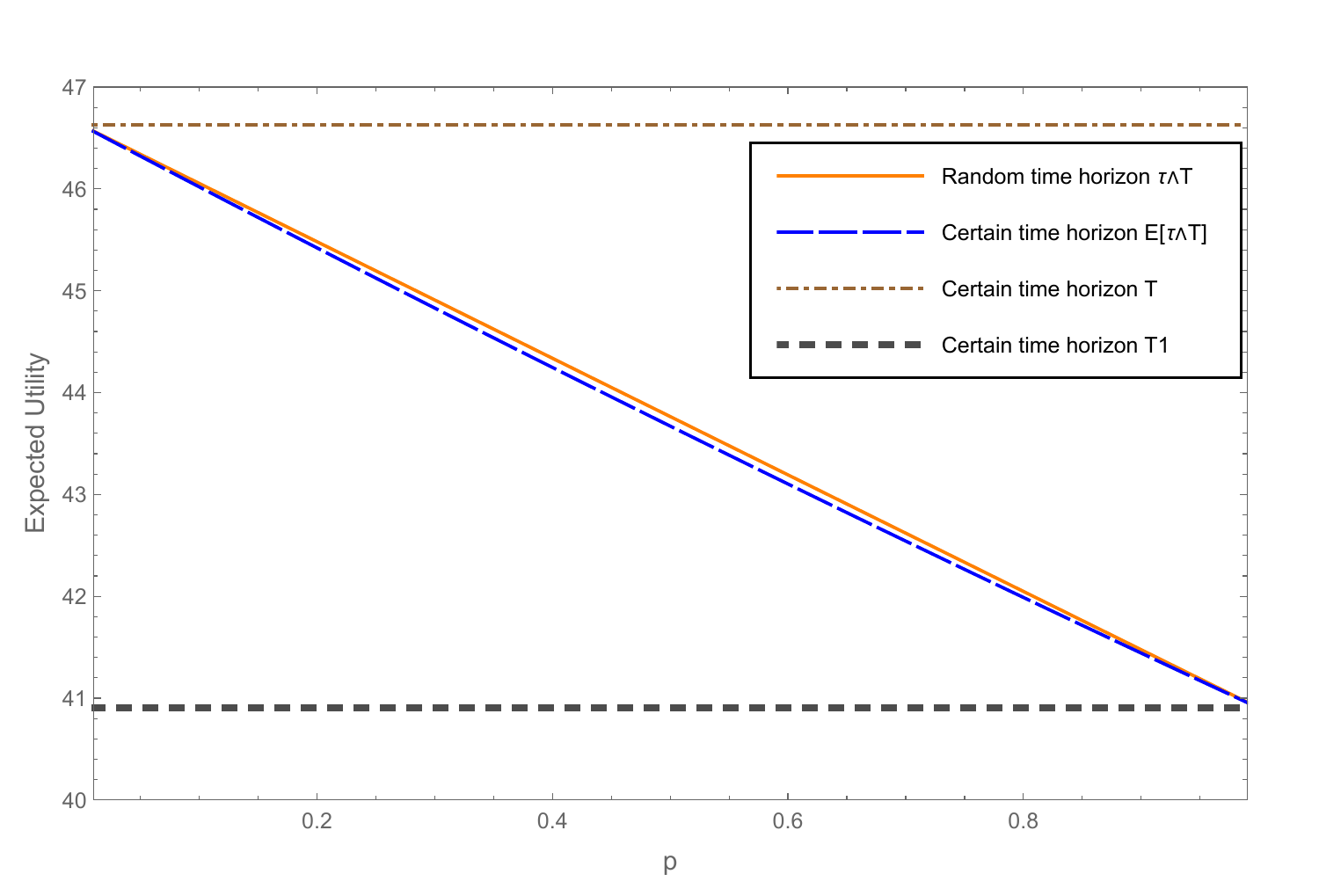}
		\caption{\small Concave problem }
       \end{subfigure}%
      \caption{{\small Impact of $p=\PP(\tau=T_1=T/2)$ on the expected utility.}}
		\label{Fig.6}
		 \end{figure}
		\end{center}
		\vspace{-2mm}
		Lastly, let $\nu_{T_i}:=\xi_{T_i}^{-1}U'(P^{{*}}_{T_i})$, $i=1,2$ where $U'$ is the right-hand derivative of $U$.   
We now want to numerically verify that as shown in Theorem \ref{thm:prop2}, the weighted multiplier $ p_1 \nu_{T_1}+p_2\nu_{T_2}$ 
is constant (a.s.) on the set $A=\{\omega:\, P^{{*}}_{T_i}(\omega)>0\}=\{\omega:\, P^{{*}}_T(\omega)>0\}$. We remark that $\{P^{{*}}_{T}=0\}$ is a non-zero set and $U$ is not differentiable at $0$. For the given parameters and for 50000 paths on the market price of risk, we obtain that $p\nu_{T_1}+(1-p)\nu_{T_2}=0.165183$ is constant on the set $A=\{P^{{*}}_{T}>0\}$, confirming the result established in Theorem \ref{thm:prop2}. Note that this weighted multiplier coincides with the multiplier of the first period. The result is consistent when different values of $p$ are considered, see Table \ref{Tab:A}. 

\section{Optimal investment with an $\cF$-stopping time}\label{Sec:stopping}
In this section we study the case where $\bar{\tau}$ is an $\cF$-stopping time taking values at $0<T_1<\cdots <T_n=T$. For simplicity, we consider again in this section the non-concave utility function $U$ defined in \eqref{eq:carpenter_utility}. We remark that the result obtained in this section can be extended to more general utilities. The optimization problem \eqref{eq:general_problem} becomes 
\begin{align}
V_{\bar{\tau}}(x,U) =  \sup_{\pi \in \Pi(0,x)} \EX \left[U(P_{T\wedge\bar{\tau}} ) \right]= \sup_{\pi \in \Pi(0,x)} \EX \left[\sum_{i=1}^{n} U(P_{T_i} ) {\bf 1}_{\bar{\tau} = T_i} \right].
\label{eq:manyTprobstopping}
\end{align}
Recall the generalized inverse marginal utility $I$ defined by \eqref{eq:Iinverse}. The function $U^c$ is not differentiable everywhere but the superdifferential $\partial{U}^c $ may be identified with the set-valued function
\begin{equation}
\label{eq:superdiff}
\partial {U}^c (x) := \begin{cases}
[U'(\hat{x}(B)),\infty) & \text{ for } x = 0, \\
\{ U'(\hat{x}(B))\} &\text{ for } 0 < x \leq \hat{x}(B), \\
\{ U'(x) \} & \text{ for } x > \hat{x}(B).
\end{cases}
\end{equation}

\begin{proposition} \label{thm:stoppingtime} Assume that $\bar{\tau}$ is an $\cF$-stopping time taking values $0<T_1<\cdots <T_n=T$. 
Suppose furthermore that there is an {adapted process} $\nu\ge 0$ with $\nu_0=U'(x)$ such that the process $\xi P^{x,\pi^{(n)}(x,I(\nu_{\bar{\tau}\wedge T} \xi_{\bar{\tau}\wedge T}))}$ generated by $(x,I(\nu_{\bar{\tau}\wedge T} \xi_{\bar{\tau}\wedge T}))$ is a martingale and {$ \nu_{\bar{\tau}\wedge T}=\sum_{i=1}^{n} \nu_{T_i}{\bf 1}_{\bar{\tau} = T_i}$ is a constant}. Then, $\xi P^{x,\pi^{(n)}(x,I(\nu_{\bar{\tau}\wedge T} \xi_{\bar{\tau}\wedge T}))}$ solves the optimization problem \eqref{eq:manyTprobstopping}.
\end{proposition} 
\proof 
Let $ \sum_{i=1}^{n} \nu_{T_i}\mathbbm{1}_{\bar{\tau} = T_i}=y$ which is a constant by assumption. Let us first show that for any $Y=(x,Y_{\bar{\tau}\wedge T})\in C_{\bar{\tau}}(x)$ we have 
\begin{equation}
\EX[\sum_{i=1}^{n} \nu_{T_i}  \xi_{T_i} Y_{T_i} \mathbbm{1}_{\bar{\tau} = T_i}]\le xy, \quad \forall \,Y=(x,Y_{\bar{\tau}\wedge T})\in C_{\bar{\tau}}(x).
\label{eq:dualityineqstopping}
\end{equation}
Indeed, by construction the process $\xi Y$ is a local martingale. Denoting the localizing sequence of stopping times $\le T$ of this local martingale by $(\bar{\tau}_n)$, we have
$$
\EX[\mathbbm{1}_{\bar{\tau} = T_i}\nu_{T_i}\xi_{\bar{\tau}_{n} \wedge T} Y_{\bar{\tau}_{n} \wedge T}\vert \cF_{T_i}]\ge \mathbbm{1}_{\bar{\tau} = T_i}\mathbbm{1}_{\bar{\tau}_n\ge T_i}\nu_{T_i}\xi_{T_i} Y_{T_i}.
$$
Using Fubini's Theorem we obtain
\begin{align*}
\EX[\sum_{i=1}^{n}\mathbbm{1}_{\bar{\tau}_n\ge T_i} \mathbbm{1}_{\bar{\tau} = T_i}\nu_{T_i}\xi_{T_i}Y_{ T_i}]
&\le\EX\bigg[\sum_{i=1}^{n} \EX\big[\mathbbm{1}_{\bar{\tau} = T_i}\nu_{T_i}\xi_{\bar{\tau}_{n} \wedge T}Y_{\bar{\tau}_{n}\wedge T}\vert \cF_{T_i}\bigg]\\
&=\EX\bigg[\xi_{\bar{\tau}_{n} \wedge T} Y_{\bar{\tau}_{n} \wedge T} \sum_{i=1}^{n} \mathbbm{1}_{\bar{\tau} = T_i}\nu_{T_i}\bigg] 
=y\EX\bigg[\xi_{\bar{\tau}_{n} \wedge T} Y_{\bar{\tau}_{n} \wedge T}  \bigg] =xy.
\end{align*}
The conclusion follows by passing to the limit and Fatou's lemma.


Note that the $Z_\zs{\bar{\tau} \wedge T}:=x^{-1}\xi_\zs{\bar{\tau} \wedge T} P^{x,\pi^{(n)}(x,I(\nu_{\bar{\tau} \wedge T)}\xi_{\bar{\tau} \wedge T)})}$ defines a density process of a probability measure $\QQ^\nu<<\PP$ as it is a martingale with initial value equal to 1. By construction, $Z_{T_i}=x^{-1}\xi_{T_i}I(\nu_{T_i} \xi_{T_i})$. Therefore, by Bayes formula we obtain
\begin{equation}
\EX[\sum_{i=1}^{n} {\bf 1}_{\bar{\tau} = T_i} \nu_{T_i} \xi_{T_i}I(\nu_{T_i} \xi_{T_i})]= x\EX^{\QQ^\nu}[\sum_{i=1}^{n} {\bf 1}_{\bar{\tau} = T_i} \nu_{T_i} ]=xy.
\label{eq:budget}
\end{equation}
For any admissible $Y=(x_0,Y_{T_1},\cdots, Y_{T_n})\in C_{\bar{\tau}}(x)$ we have by \eqref{eq:budget} that
\begin{align*}
\EX \left[\sum_{i=1}^{n} {\bf 1}_{\bar{\tau} = T_i}U(I(\nu_{T_i} \xi_{T_i})) \right]&=\EX \left[\sum_{i=1}^{n} {\bf 1}_{\bar{\tau} = T_i} U(I(\nu_{T_i} \xi_{T_i})) \right]- x\EX^{\QQ^\nu}[\sum_{i=1}^{n} {\bf 1}_{\bar{\tau} = T_i} \nu_{T_i} ]+xy\\
&=\EX \left[\sum_{i=1}^{n} {\bf 1}_{\bar{\tau} = T_i} \bigg(U(I(\nu_{T_i} \xi_{T_i})) -\nu_{T_i} \xi_{T_i}I(\nu_{T_i} \xi_{T_i})\bigg)\right ]+xy
\\
&=\EX \left[\sum_{i=1}^{n} {\bf 1}_{\bar{\tau} = T_i} \sup_{X\ge 0}\bigg(U(X) -\nu_{T_i} \xi_{T_i}X\bigg)\right ]+xy\\
&\ge\EX \left[\sum_{i=1}^{n} {\bf 1}_{\bar{\tau} = T_i} \bigg(U(Y_{T_i}) -\nu_{T_i} \xi_{T_i}Y_{T_i}\bigg)\right ]+xy
\ge \EX \left[\sum_{i=1}^{n} {\bf 1}_{\bar{\tau} = T_i} U(Y_{T_i})\right ],
\end{align*}
where we have used \eqref{eq:dualityineqstopping} in the last step. This implies the optimality of the process $ P^{x,\pi^{(n)}(x,I(\nu_{\bar{\tau}\wedge T} \xi_{\bar{\tau}\wedge T}))}$ generated by $(x,I(\nu_{\bar{\tau}\wedge T} \xi_{\bar{\tau}\wedge T}))$. \endproof

We now aim to solve the non-concave optimization problem when $\bar{\tau}$ is an $\cF$-stopping time, namely 
\begin{align}
V_{\bar{\tau}}(x,u) =  \sup_{\pi \in \Pi(0,x)} \EX \left[U(P_{T\wedge\bar{\tau}} ) \right]= \sup_{\pi \in \Pi(0,x)} \EX \left[\sum_{i=1}^{n} U(P_{T_i} ){\bf 1}_{\bar{\tau} = T_i} \right],
\label{eq:manyTprobstopping2}
\end{align}
where $U$ is the non-concave utility function defined by \eqref{eq:carpenter_utility}. In particular, by applying Proposition \ref{thm:stoppingtime} we prove below that Problem \eqref{eq:manyTprobstopping2} can be solved by concavification arguments and the optimal wealth can be characterized by $I(\nu_{\bar{\tau}\wedge t} \xi_{\bar{\tau}\wedge t})$, where $I$ is the generalized inverse marginal utility function defined by \eqref{eq:def_i} and $\nu$ is an adapted process. We need the following integrability condition.

\vspace{2mm}
\noindent{\bf Condition (C)}: For any $y>0$, $\EX \left[\xi_{\bar{\tau} \wedge T} I(y\xi_{\bar{\tau} \wedge T})\right]<\infty$.
\vspace{2mm}

Below we show that under condition (C) and the assumption that the stopping time adapted to the financial market filtration, it is possible to construct an adapted process $\nu$ such that the process generated by the stopped $n$-tuple $(x,I(\nu_\zs{\bar{\tau}\wedge T} \xi_\zs{\bar{\tau}\wedge T}))$ is a martingale and $\nu_\zs{\bar{\tau}\wedge T}=\sum_{i=1}^{n}{\bf 1}_{\bar{\tau} = T_i} \nu_{T_i}$ is a constant. The result is summarized in the following proposition.

	\begin{proposition} [Non-concave problem with a stopping time horizon]Assume that $\bar{\tau}$ is an $\cF$-stopping time taking values at $0<T_1<\cdots <T_n=T$ and Condition (C) holds.  Then, there exists an $\cF$-adapted process $\nu$ such that the optimal wealth of Problem \eqref{eq:manyTprobstopping2} is given by 
$$P^*_{\bar{\tau}\wedge T_i}:=I(\nu_{\bar{\tau}\wedge T_i} \xi_{\bar{\tau}\wedge T_i}),\quad 1\le i\le n$$ 
and {$\sum_{i=1}^nv_{T_i}{\bf 1}_{\bar{\tau} = T_i}=y^*$ is a constant} satisfying $\EX\left[\xi_{\bar{\tau} \wedge T}I(y^*\xi_{\bar{\tau} \wedge T})\right]=x.$  

 \end{proposition}

\proof
Consider the mapping $y\longmapsto \EX \left[\xi_{\bar{\tau} \wedge T} I(y\xi_{\bar{\tau} \wedge T})\right]=f(y)$ defined for $y\in (0,\infty)$ by Condition (C). Since the market price density $\xi$ is atomless, $f$ is continuous on $(0,\infty)$. Moreover, by Fatou's lemma, \eqref{eq:def_i} and Inada's condition of the power utility function $U$ we obtain $\lim_{y\to 0}f(x)=\infty$ and $\lim_{y\to \infty}f(x)=0$. Therefore, there exists $y^*\in (0,\infty)$ such that $\EX\left[\xi_{\bar{\tau} \wedge T}I(y^*\xi_{\bar{\tau} \wedge T})\right]=f(y^*)=x$. Define for $0\le t\le T$, $\zeta_t:=I(y^*\xi_{t})$ and 
$$
\nu_{t}\in\frac{1}{\xi_{t}}\partial U^c\left(\EX\left[\xi_{\bar{\tau} \wedge T}\xi^{-1}_{\bar{\tau}\wedge t}\zeta_{\bar{\tau} \wedge T}\vert \cF_{\bar{\tau}\wedge t}\right]\right),
$$
where the superdifferential  is defined by \eqref{eq:superdiff}. Note that since the conditional expectation process $\EX\left[\xi_{\bar{\tau} \wedge T}\xi^{-1}_{\bar{\tau}\wedge t}\zeta_{\bar{\tau} \wedge T}\vert \cF_{\bar{\tau}\wedge t}\right]=\EX\left[\xi_{\bar{\tau} \wedge T}\xi^{-1}_{\bar{\tau}\wedge t}\zeta_{\bar{\tau} \wedge T}\vert \xi_{\tau \wedge T}\right]>0$ is a non-negative martingale with initial value $x>0$, almost surely $\partial U^c$ corresponds to $U'$ and is invertible. 
Thus, by construction we obtain $P^*_{\bar{\tau}\wedge T}=I(\nu_{\bar{\tau}\wedge T} \xi_{\bar{\tau}\wedge T})$ with $y^*=\nu_{\bar{\tau}\wedge T}=\sum_{i=1}^nv_{T_i}{\bf 1}_{\bar{\tau} = T_i}$ being a constant and the stopped-tuple $(x,I(\nu_{\bar{\tau}\wedge T} \xi_{\bar{\tau}\wedge T}))$ is a martingale with starting value $x$. Hence, it is an optimal solution to \eqref{eq:manyTprobstopping2}. \endproof

The following is aligned with Proposition 3.3 in \cite{BouchardPham} when $\bar{\tau}$ is an $\cF$-stopping time for stricly concave utility function $U$. 

\begin{corollary} [Concave problem with a stopping time horizon] Assume $U$ that is a strictly concave utility for which Condition (C) holds and $\bar{\tau}$ is an $\cF$-stopping time taking values at $0<T_1<\cdots <T_n=T$.  For any $x>0$, there exists $y^*>0$ such that $\EX \left[\xi_{\bar{\tau} \wedge T} I(y^*\xi_{\bar{\tau} \wedge T})\right]=x$. Moreover, there exists an adapted process $\nu$ such that the optimal wealth of Problem \eqref{eq:manyTprobstopping2} is given by $P^*_{\bar{\tau}\wedge T_i}:=I(\nu_{\bar{\tau}\wedge T_i}),\quad 1\le i\le n$, 
and $\nu_{\bar{\tau}\wedge T}=\sum_{i=1}^n\nu_{T_i}{\bf 1}_{\bar{\tau} = T_i}=y^*$ is a constant.
\end{corollary}

\section{Conclusion}\label{Se:conclusion}
We studied a non-concave optimal investment with a random time horizon in a complete financial market setting. We established a necessary and sufficient condition for the optimality in this case for general utility functions with a random time horizon.  When $\tau$ is independent of the financial risk, we showed that a direct concavification approach cannot be applied and suggest a recursive procedure based on the dynamic programming principle. We illustrated our finding by carrying out a multiple period numerical analysis for the non-concave option compensation problem with random time horizon. 
We numerically show that due to concavification, the distribution of the wealth at exiting times of the non-concave optimization problems is right-skewed with a long right tail, indicating that the investor can expect frequent small losses and a few large gains from the investment. Under the premature  exiting risk, the wealth at an exiting time exhibits a bimodal distribution with peaks of different heights due to the concavification procedure and  whereas the exiting time $\tau$ distribution has significant impact on the  amplitude between the two modes. 

Our work leaves several interesting directions for future work. for instance, it would be interesting to look at the case when the time horizon is correlated with the financial market information, or to investigate the problem in a general incomplete financial market like in \cite{BouchardPham}. Furthermore, our non-concave framework with random horizon might serve as an attempt to extend the results for contract design problems of term-life insurance or insurance contracts with surplus participation  \cite{Chen,Stadje} to an uncertain time horizon setting. We leave this for future work.  

\vspace{0.2cm}

{\raggedleft{\textbf{\large{Acknowledgments}}}}: {Thai Nguyen acknowledges the support of the Natural Sciences and Engineering Research Council of Canada [RGPIN-2021-02594]. 
}

\section{Appendix}
The following result can be shown directly using the lognormal distribution of $\xi$:
\begin{lemma} \label{lemma:Thais_lemma1}
Let $q \in \ree$. With $f$ defined in Proposition \ref{pro:concavepower1} it holds for $0 \leq t \leq T$ that
\begin{align}
\EX \left[ \xi_T^q | \mathcal{F}_t \right] = \EX \left[ \left(\frac{\xi_T}{\xi_t} \right)^q \bigg| \mathcal{F}_t \right] \xi_t^{q} = f(q,t,T) ~\xi_t^{q}.
\end{align}
\end{lemma}
The next result provides a generalization of Lemma \ref{lemma:Thais_lemma1} when the market parameters are constant. 
\begin{lemma} \label{lemma:Thais_lemma2}
Let $q \in \ree$, $0 \leq t < T$ and let $\lambda$ be a positive constant. With $\Phi$ the cdf of the standard normal distribution and $d$ defined in (\ref{eq:d}) it holds that 
\begin{align} \label{eq:Thais_lemma2}
\EX \left[ \xi_T^q {\bf 1}_{\lambda \xi_T \leq U'(\hat{x}(B))} | \mathcal{F}_t \right] = ~\xi_t^{q} f(q,t,T)\Phi \left( d(q,t,T,\lambda \xi_t) \right).
\end{align}
\end{lemma}

\begin{lemma}\label{Le:conf}
Let $U$, $V$ be continuous, increasing functions in $[0,\infty)$. Let $(a,b)\subset\{U<U^c\}$ be an open interval in the concavification region of $U$. Assume that there exists $x_0\in[a,b)$ at which $U+V$ coincides with the affine line 
$$
g(x):=U(a)+V(a)+\frac {(U(b)+V(b))-(U(a)+V(a))}{b-a}(x-a)
$$
and the right derivative of the sum $U+V$ exists and
\begin{equation}
U'(x_0^+)+V'(x_0^+)>\frac {(U(b)+V(b))-(U(a)+V(b))}{b-a}.
\label{eq:UV}
\end{equation}
Then, the interval $(a,b)$ cannot be a concavification set of the sum $U+V$, i.e. there exists an open interval $(a',b')\subset(a,b)$ such that $U(x)+V(x)=(U+V)^c(x)$ for all $x\in(a',b')$.
\end{lemma}

\proof We have $U(x_0)+V(x_0)=g(x_0)$. By continuity and \eqref{eq:UV} it can be seen that $U+V>g(x)$ in  a right-hand neighbourhood of $x_0$, which implies that the affine line $g$ is not the concave hull of $U+V$ on the whole interval $(a,b)$.
\endproof


\begin{thebibliography}{9}  

\bibitem{filtration} 
\textsc{Aksamit, A. and Jeanblanc, M.}: Enlargement of filtration with finance in view. \textit{Switzerland: Springer,} (2017).


\bibitem {AumannPerles} \textsc{Aumann, R. J. and Perles, M.}:\ A variational problem arising in economics, \textit{Journal of Mathematical Analysis and Applications} \textbf{11}, (1965), 488-503.



\bibitem {BasakMakarov} \textsc{Basak, S. and Makarov, D.}:\ Strategic asset allocation in money management, \textit{The Journal of Finance} \textbf{69}(1), (2014), 179-217.


\bibitem{Bensoussan} \textsc{Bensoussan, A. Cadenillas, and H. K. Koo.}:\ Entrepreneurial decisions on effort and project
with a nonconcave objective function. \textit{Mathematics of Operations Research} \textbf{40}(4), (2015) 902–914.




\bibitem {Biagini_survey} \textsc{Biagini, S.}:\ Expected Utility Maximization: Duality Methods, \textit{Encyclopedia of Quantitative Finance}, (2010).




\bibitem{Biane2011}
\textsc{Bian, B., Miao, S., and Zheng, H.}:\ Smooth value functions for a class of nonsmooth utility maximization problems.  \textit{SIAM Journal on Financial Mathematics}, \textbf{2}(1), (2011), 727-747.

\bibitem{Biane2019}
\textsc{Bian, B., Chen, X., and Xu, Z. Q.}:\  Utility maximization under trading constraints with discontinuous utility.  \textit{SIAM Journal on Financial Mathematics}, \textbf{10}(1), (2019), 243-260.

\bibitem {BichuchSturm} \textsc{Bichuch, M. and Sturm, S.}:\ Portfolio optimization under convex incentive schemes, \textit{Finance and Stochastics} \textbf{18}(4), (2014), 873-915.



\bibitem {Blanchet2005} \textsc{Blanchet-Scalliet, C., El Karoui, N. and Martellini, L.}:\ Dynamic asset pricing theory with uncertain time-horizon, \textit{Journal of Economic Dynamics and Control} \textbf{29}(10), (2005), 1737-1764.

\bibitem {Blanchet} \textsc{Blanchet-Scalliet, C., El Karoui, N., Jeanblanc, M. and Martellini, L.}:\ Optimal investment decisions when time-horizon is uncertain, \textit{Journal of Mathematical Economics} \textbf{44}(11), (2008), 1100-1113.


\bibitem {BouchardPham} \textsc{Bouchard, B. and Pham, H.}:\ Wealth-path dependent utility maximization in incomplete markets, \textit{Finance and Stochastics} \textbf{8}(4), (2004), 579-603.






\bibitem {CarassusPham} \textsc{Carassus, L. and Pham, H.}: \ Portfolio optimization for piecewise concave criteria functions, \textit{The 8th Workshop on Stochastic Numerics}, (2009).

\bibitem {Carpenter} \textsc{Carpenter, J. N.}:\ Does option compensation increase managerial risk appetite?, \textit{The Journal of Finance} \textbf{55}(5), (2000), 2311-2331.

\bibitem{Chen} \textsc{Chen, A., Hieber, P. and Nguyen, T.}:\ Constrained non-concave utility maximization: an application to life insurance contracts with guarantees, \textit{European Journal of Operational Research} \textbf{273}(3), (2019), 1119-1135.



\bibitem {Dai}  \textsc{Dai, M., Kou, S., Qian, S. and Wan, X.}:\ Nonconcave Utility Maximization without the Concavification Principle, (2019). Available at SSRN: https://ssrn.com/abstract=3422276.



\bibitem{online}
\textsc{Dehm, C., Nguyen, N., and Stadje, M.} Non-concave expected utility optimization with uncertain time horizon. {\textit arXiv preprint}: https://arxiv.org/pdf/2005.13831.pdf (2021).



























\bibitem{Karatzas} \textsc{Karatzas, I., Lehoczky, J. P., Shreve, S. E., and Xu, G. L.:} Martingale and duality methods
	for utility maximization in an incomplete market, \textit{SIAM Journal on Control and Optimization} \textbf{29}(3), (1991),
	702-730.
	




\bibitem {Larsen} \textsc{Larsen, K.}:\ Optimal portfolio delegation when parties have different coefficients of risk aversion, \textit{Quantitative Finance} \textbf{5}(5), (2005), 503-512.



\bibitem {Merton1971} \textsc{Merton, R.C.}:\ Optimal consumption and portfolio rules in continuous time, \textit{Journal of Economic Theory} \textbf{3}, (1971), 373-413.



\bibitem{Stadje} \textsc{Nguyen, T. and Stadje, M.}:\ Nonconcave optimal investment with Value-at-Risk constraint: an application to life insurance contracts, \textit{SIAM Journal on Control and Optimization}, \textbf{58}(2), (2020), 895-936.



\bibitem {Pham_enlargement} \textsc{Pham, H.}:\ Stochastic control under progressive enlargement of filtrations and applications to multiple defaults risk management, \textit{Stochastic Processes and their Applications} \textbf{120}(9), (2010), 1795-1820.

\bibitem {Reichlin} \textsc{Reichlin, C.}:\ Utility maximization with a given pricing measure when the utility is not necessarily concave, \textit{Mathematics and Financial Economics} \textbf{7}(4), (2013), 531-556.

\bibitem {Richard} \textsc{Richard, S. F.}:\ Optimal consumption, portfolio and life insurance rules for an uncertain lived individual in a continuous time model, \textit{Journal of Financial Economics} \textbf{2}(2), (1975), 187-203.

\bibitem {Rieger2012} \textsc{Rieger, M. O.}:\ Optimal financial investments for non-concave utility functions, \textit{Economics Letters}, \textbf{114}(3), (2012), 239-240.



\bibitem{ross2004}\textsc{Ross S.~A.}: \ Compensation, incentives, and the duality of risk aversion and riskiness, \textit{Journal of Finance}, \textbf{59} (1), (2004), 207-225.






\bibitem{Wong} \textsc{Wong, K. C., Yam, S. C. P., and Zheng, H.:} Utility-deviation-risk portfolio selection. \textit{SIAM Journal on Control and Optimization} \textbf{55}(3), (2017), 1819-1861.

\bibitem {Yaari} \textsc{Yaari, M. E.}:\ Uncertain lifetime, life insurance, and the theory of the consumer, \textit{The Review of Economic Studies} \textbf{32}(2), (1965), 137-150.





\end{thebibliography}
\end{document}